\documentclass[twocolumn]{aastex631}

\newcommand{\msun}{M_\odot}
\newcommand{\mdot}{\dot{m}}

\usepackage{xcolor}

\usepackage{amsmath}

\let\tablenum\relax
\let\tablewidth\relax

\usepackage{siunitx}
\usepackage{graphicx}
\usepackage{booktabs}
\graphicspath{figures}
\usepackage{tabularray}
\usepackage{soul}

\begin{document}

\title{Simulation-Based Prediction of Black Hole Fe K$\alpha$ Line Profiles}

\author[0000-0003-0685-3525]{Chris Nagele}
\affiliation{Department of Physics and Astronomy\\
Johns Hopkins University\\
Baltimore MD 21218}

\author[0000-0002-2995-7717]{Julian H. Krolik}
\affiliation{Department of Physics and Astronomy\\
Johns Hopkins University\\
Baltimore MD 21218}

\author[0000-0002-8676-425X]{Brooks E. Kinch}
\affiliation{Department of Mechanical Engineering and Applied Mechanics\\
University of Pennsylvania\\
Philadelphia PA 19104}

\author[0000-0002-2942-8399]{Jeremy D. Schnittman }
\affiliation{Gravitational Astrophysics Lab, NASA Goddard Space Flight Center, Greenbelt MD 20771}

\email{chrisnagele.astro@gmail.com}

\begin{abstract}

One of the most useful spectral diagnostics of accreting black hole
systems is the Fe K$\alpha$ fluorescence line. Detected in many systems,
it is often used to estimate the black hole spin, as its breadth is
attributed to relativistic kinematics near the spin-dependent innermost
stable circular orbit (ISCO).  In a companion paper, we showed how
continuum spectra emitted by accreting black holes can be derived from
snapshots of general relativistic magnetohydrodynamics simulations by
combining radiation transfer solutions for the disk body and the
corona.  In this paper, we focus on the Fe K$\alpha$ line, solving its
transfer problem on the basis of local ionization and thermal balance. 
Its equivalent width is $\sim 25-225$ eV, depending mainly on viewing
angle, for an accretion rate of 1$\%$ Eddington.  Contrary to common assumptions, the illuminating X-ray spectrum
and ionization parameter $\xi$ can be strong functions of radius; e.g. $\xi \propto r^{-1.5}$ in this simulation.   Consequently, the region of the
disk near the ISCO is completely ionized and contributes almost no Fe
K$\alpha$ photons; most of the flux is made at radii $\gtrsim 10 r_g$. 
The lines are broadened by a combination of relativistic Doppler shifts,
Compton broadening in the disk atmosphere, and the differing line
energies emitted by different Fe ions.
These new mechanisms expand the parameter space of acceptable models, including the possibility of broad line profiles without large black hole spin; physical trends revealed by the simulations can refocus fitting efforts on the most relevant sections of the parameter space.

\end{abstract}


\section{Introduction} \label{sec:intro}

Broad iron emission lines are a common feature of observed X-ray spectra of accreting black holes \citep{Reynolds2003PhR...377..389R}. After broad iron K$\alpha$ lines were observed in both stellar mass \citep{Fabian1989MNRAS.238..729F} and supermassive black hole (BH) sources \citep{Tanaka1995Natur.375..659T}, a consensus emerged that the broadness of these emission lines was due in the main to extreme relativistic effects near the black hole \citep[][ and references therein]{Reynolds2003PhR...377..389R}. This birthed the field of X-ray reflection spectroscopy, whereby parameters of the black hole could be inferred by the shape of the iron line. Foremost among these parameters was the black hole spin, the idea being that if the accretion disk (off which X-rays 'reflect', hence X-ray reflection) terminated abruptly at the innermost stable circular orbit (ISCO), then the location of the ISCO could be determined from the line profile. The ISCO has a monotonic relation with the black hole spin, thus allowing a spin measurement to be made. There are, of course, other parameters involved (density, temperature, ionization parameter, inclination angle) and modern efforts utilize 'phenomenological models' to scan the large parameter space using statistical methods \citep{Ross2005MNRAS.358..211R,Garcia2010ApJ...718..695G,Garcia2014ApJ...782...76G,Huang2025arXiv251212728H}. They generally infer black hole spin parameters $a>0.9$ \citep[e.g.][]{Jiang2019MNRAS.489.3436J,Draghis2025ApJ...989..227D}, slightly higher than spins inferred by X-ray continuum fitting \citep[e.g.][]{Shafee2006ApJ...636L.113S}, and much higher than  spins inferred from the population of compact object coalescences \citep{Abbott2023PhRvX..13a1048A}.

Despite the ubiquity of X-ray reflection spectroscopic black hole spin measurements, this program has not yet been tested by simulations based on fundamental physics. In this paper, we build upon previous work \citep{Schnittman_2013,Kinch_2016,Kinch_2019,Kinch_2021,Liu+2025,Nagele2026arXiv260103349N} in order to take the next step in performing such a test.
The starting point for this work is data snapshots showing 3D maps of density, velocity, and heating rate calculated by general relativistic (GR) magnetohydrodynamic (MHD) simulations.  The simulations on which the present work is based are wholly specified by only a few parameters: black hole spin, accretion rate in Eddington units, the disk's target scale-height, and the geometry of the initial magnetic field.  Their results can be scaled after the fact to reflect the black hole mass.  Because the physics of matter-photon interactions in this regime is well-known, the snapshot data completely determine the output spectrum, but they do so via solutions to the equations of ionization balance, thermal balance, and radiation transfer.

Previous work has generally assumed that all hard X-rays are generated in a compact volume, all line photons are generated inside the photosphere of the disk body, and all other regions of the system are transparent.  These assumptions permit treating the production of hard X-rays as independent from the ``reflection" spectrum they generate in the disk.  However, the simulation data yields continuous distributions of density and dissipation.   Consequently, nowhere within the accretion system is free from either opacity or dissipative heating, and all of it must be brought into an internally self-consistent state in order to predict the spectrum.  Thermal photons from the disk photosphere are upscattered throughout the surrounding low-density ``corona", in which there are large variations in temperature; some of these photons return with boosted energy to ionize gas in the disk; and the spectrum of seed photons leaving the disk photosphere is no longer strictly thermal.

To be more specific, we divide the global problem into two parts.  Outside the disk photosphere,
we use the relativistic Monte Carlo code \texttt{Pandurata} \citep{Schnittman_2013b} to solve for local thermal balance throughout the corona and the spectral intensity at both the disk surface and the exterior boundary of the simulation. The flux traveling out of the disk and into the corona sets the boundary condition for \texttt{Pandurata}.
Inside the disk body, we solve a large number of local 1D radiation transfer problems in order to find the spectrum emitted from each patch of the disk's photosphere.  The local opacities and emissivities within the associated columns are computed from the results of local thermal and ionization balance solutions.
Not only does this calculation yield the correct seed photon spectrum for \texttt{Pandurata}, these spectra self-consistently include all atomic emission lines and absorption troughs.
We iterate between the corona and disk-body solutions in order to guarantee full self-consistency.

\citet{Nagele2026arXiv260103349N} performed 
a black hole mass survey using these methods, and found that much of the observed spectral behavior of accreting stellar mass black holes and supermassive black holes can be explained by \textit{the same 3D-GRMHD simulations and radiation physics}. In this paper, we extend this approach to emission lines by performing radiation transfer calculations on an energy grid with much higher resolution.  
We find that for all black hole masses with accretion rates of $1\%$ Eddington, Fe K$\alpha$ line have equivalent widths in the range $25-225$ eV and have profiles closely resembling many observed iron lines. This is despite almost no iron line emission from the accretion disk near the ISCO, a region which is fully ionized due to X-ray irradiation. In Sec. \ref{sec:methods} we outline our numerical methods. In Sec. \ref{results:disk_flux}, we describe the ionization state of the accretion disk and the radiation in the fluid frame. In Sec. \ref{results:obs_flux} we describe the radiation in the observer frame. We end with a discussion and conclusions in Secs. \ref{sec:discussion}, \ref{sec:conclusions}.

\section{Methods} \label{sec:methods}
\subsection{\texttt{HARM3D}: the underlying GRMHD simulation}  \label{sec:methods_harm}

Both of the GRMHD simulation snapshots used in this paper and in \citet{Nagele2026arXiv260103349N} are \texttt{HARM3D} snapshots. \texttt{HARM3D} solves the GRMHD equations in flux conservative form in Modified Kerr-Schild coordinates \citep{Gammie2003ApJ...589..444G}, and is suitable for the simulation of accretion onto a spinning black hole from the event horizon out to $\sim 100 \;r_g$ (where $r_g = GM/c^2$ is the gravitational radius). In this paper, the spin parameter is $a=J/M=0.9$ (so that the ISCO is at $\simeq 2.3\; r_g$) and the simulation domain extends to $r=70\;r_g$.

We model black holes with eight different masses, from $10\;\msun$ to $10^8\;\msun$ and two different accretion rates in Eddington units, $\dot m = 0.01, 0.1$, as in \citet{Nagele2026arXiv260103349N}.  Because the cooling function in the corona depends on $\dot m$, it must be specified for each simulation.  However, final conversion to physical units, which carries the dependence on black hole mass $M$, can be done after the fact through relations like those for density and luminosity:
\begin{align}
\rho_{\text {cgs }} r_g &= \rho_{\text {code }} \frac{4 \pi }{\kappa_T} \frac{\mdot / \eta}{\mdot_{\text {code }}} \label{Eq:rho_cgs}\\
\mathcal{L}_{\text {cgs }} &=\mathcal{L}_{\text {code }} \frac{4 \pi c^3}{\kappa r_g^2} \frac{\mdot / \eta}{\mdot_{\text {code }}}.
\end{align}
Here $r_g=GM/c^2$ plays the role of the code-unit for length; it can be freely reinterpreted in terms of physical units by choosing $M$ after the simulation has been run.  In addition, $\kappa_T = 0.4$~{cm$^2$~g$^{-1}$ is the electron scattering opacity and $\eta =0.1558$ is the Novikov-Thorne efficiency for an $a = 0.9$ black hole.}

\texttt{HARM3D} uses two types of cooling functions to approximate radiative losses from the system. The first is the target entropy cooling function which cools material towards a targeted entropy, and hence, cools the disk towards a specific value, e.g. H/R $\approx0.05$ \citep{Noble_2009}. We use this cooling function in the optically thick accretion disk. The second type of cooling function is a zeroth order inverse Compton cooling function \citep{Kinch_2020}. This cooling function estimates the radiation energy density in the optically thin corona originating from the disk photosphere, and then computes the rate at which the hot coronal gas will cool via inverse Compton scattering. 

\begin{figure*}
\includegraphics[width=\linewidth]{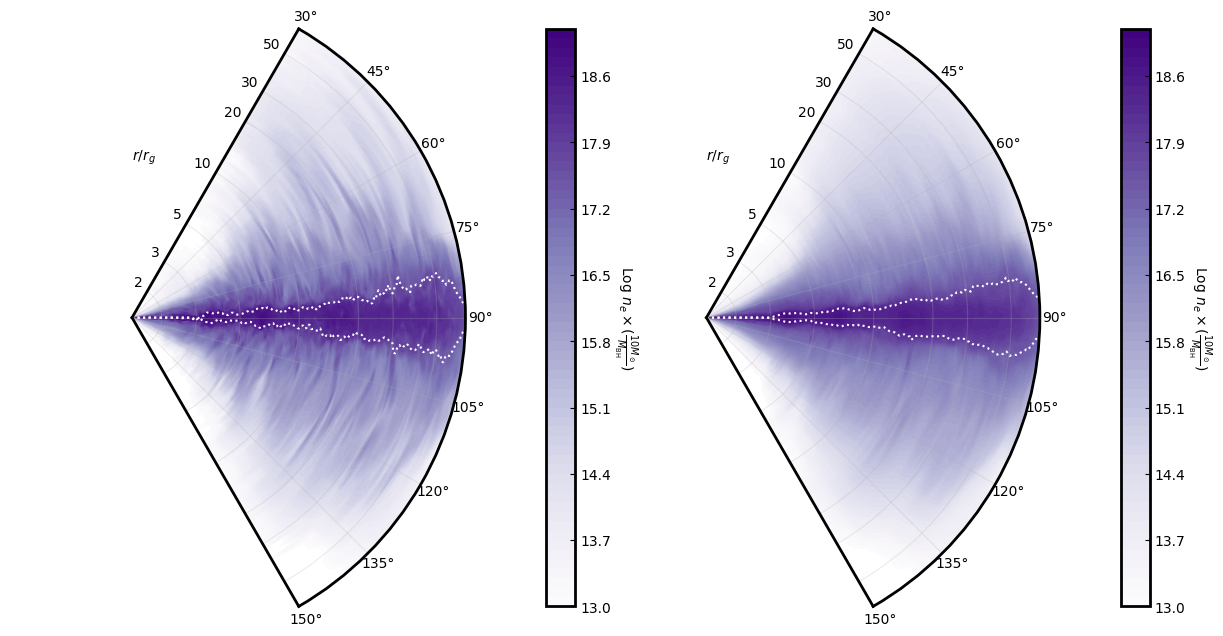}
    \caption{Electron density on a logarithmic scale in a $\mdot = 0.01$ snapshot for an azimuthal slice (left) and azimuthally averaged (right) The units are for the $M=10\;\msun$ model; the density scales with black hole mass as $n_e\propto M^{-1}$ (Eq.~\ref{Eq:rho_cgs}). The white dotted line shows the $\tau=1$ scattering photosphere in both panels. }
    \label{fig:dens}
\end{figure*}

Figure~\ref{fig:dens} shows the density in cgs units appropriate to the $M=10\;\msun$ model for one of our $\mdot=0.01$ snapshots.
As it demonstrates, the density varies quite strongly, spanning at least four orders of magnitude.  At any given moment, there can be order of magnitude contrasts on short lengthscales.
The white dotted line in this figure marks the instantaneous Thomson scattering photosphere, found by integrating the density in the $\theta$ direction from the poles towards the midplane \citep{Schnittman_2013}. The precise location of the division between ``corona" and ``disk" is not overly important \citep{Kinch_2020,Kinch_2021}, so $\tau=1$ serves as a convenient but arbitrary boundary. 
Note that the density distribution, and therefore the coronal geometry, is completely specified by the dynamical equations and the parameters of the 3D-GRMHD simulation: black hole spin, accretion rate, target entropy and initial magnetic field configuration.

\subsection{Low resolution spectra}  \label{sec:methods_lowres}

In this subsection, we describe our method of solving for thermal balance and ionization balance in each cell of the simulation snapshot. This is done with a relatively low resolution energy grid (20 gridpoints per decade), and the results of this calculation have been described in detail in \citet{Nagele2026arXiv260103349N}. In the following subsections, we describe our method for subsequently obtaining high resolution spectra (400 gridpoints per decade).

The post-processing procedure is built upon the separation of the system into two regimes, an optically thin corona and an optically thick disk. The radiative processes relevant to these two regimes differ, and require different numerical approaches, namely Monte-Carlo for the former and Feautrier for the latter. We employ two different radiation solvers, one for each of the two regimes. 

\texttt{Pandurata} is a relativistic Monte-Carlo ray tracing code \citep{Schnittman_2013b}. The code launches photon packets (or superphotons) from the disk photosphere which separates the disk from the corona. These packets encode both spectral and polarization information. As the photon packets travel through the corona, they have a chance of scattering off of an electron (probability determined by the simulation density) or reflecting off the disk (if the photon packet impacts the photosphere). 

In the scattering case, we construct lookup tables so that the Klein-Nishina cross section of each photon in the packet scattering off of each electron in that simulation cell can be efficiently computed. We take the electron distribution to be a thermal distribution with a specified electron temperature $T_e$. Changes in $T_e$ will result in changes in the Compton power discharged in a particular cell, and it is this power which we equate to the GRMHD cooling in order to bring the system into thermal balance. 

In the case of reflection, we also reference lookup tables. These are the results of a separate Monte-Carlo calculation performed with excellent resolution of the incoming photon angle. This simulation takes into account the temperature, emissivity, and absorption discussed below.

The Parallel Transport with XSTAR code (\texttt{PTransX}) is a one dimensional radiation solver \citep{Mihalas1985JCoPh..57....1M} which utilizes the XSTAR atomic database \citep{Kallman2001ApJS..133..221K} to compute emissivity and absorption for atomic transitions \citep{Schnittman_2013,Kinch_2016,Kinch_2019,Kinch_2021}. We assume that inside the accretion disk where the gas is optically thick, the radiation changes most strongly in the vertical direction, which is the direction in which there are significant changes in optical depth. With this assumption, we reduce the numerical cost of solving for the spectral and angular intensity, and we are thus able to include the (numerically expensive) atomic transitions. In the optically thick disk, we solve not only for thermal balance, but also for ionization balance. 

Each of these two codes \texttt{Pandurata} and \texttt{PTransX} provide the radiative boundary condition for the other at the location where their domains meet, namely the disk photosphere. We begin with a \texttt{Pandurata} run where we assume the radiation from the disk to be a hardened blackbody. Then we drop this assumption and iterate between the two codes. The temperature and spectral intensity both converge reasonably quickly (2-4 cycles between the two codes). We judge the system to be converged when successive \texttt{Pandurata} spectra differ by at most $5\%$ in every energy bin. 

\subsection{High resolution spectra: atomic features}  \label{sec:methods_highres_atomic}

Everything described in the previous subsection was done with a low resolution energy grid, but the study of relativistically broadened emission lines necessitates a higher energy resolution. Thus after we have found a converged solution, we do two additional computations on a higher resolution energy grid. 

The first of these is to recompute the ionization balance within the disk, without changing the thermal balance solution (that is, the temperature). To do this, we use an energy grid which has variable spacing. If we were to use a constant dLog$\nu$ spacing, the recalculation of the ionization balance would take weeks instead of hours. We use very fine spacing (400 gridpoints per decade) in the energy space relevant to Fe K$\alpha$ (3.16-17.7 keV), less fine spacing (100 gridpoints per decade) between .01 keV and 100 keV and low resolution outside of this range.

\subsection{High resolution spectra: Feautrier solution and scattering}  \label{sec:methods_highres_reflection}

Once we have computed the high resolution atomic emissivity and absorption, we perform one final Feautrier solution inside the disk. We then take the outgoing flux, which will serve as the seed photon spectra for \texttt{Pandurata}, and interpolate it back onto a constant dLog$\nu$ grid for use in \texttt{Pandurata}. 

\texttt{Pandurata} can be run with any energy grid as long as the inverse Compton scattering tables are recalculated for that energy grid. We run one final \texttt{Pandurata} simulation with the temperature once again fixed. This run takes 400 times longer ($\sim 1100 $ core hours) than the low resolution runs and is often more expensive than all previous low resolution runs combined (including \texttt{PTransX}).

\section{Results} \label{sec:results}

We divide our results into two subsections.  The first discusses the radiation spectrum at and inside the disk photosphere (Sec. \ref{results:disk_flux}). These spectra are the results of \texttt{PTransX} calculations, and are therefore computed in the fluid frame. In the second, Sec. \ref{results:obs_flux}, we propagate the disk spectra through \texttt{Pandurata}, taking all relativistic effects into account, and present the spectra seen by a distant observer.

\subsection{Disk flux} \label{results:disk_flux}

\subsubsection{Incident flux and the ionization parameter} \label{results:disk_flux_incident}

\begin{figure*}
\includegraphics[width=\linewidth]{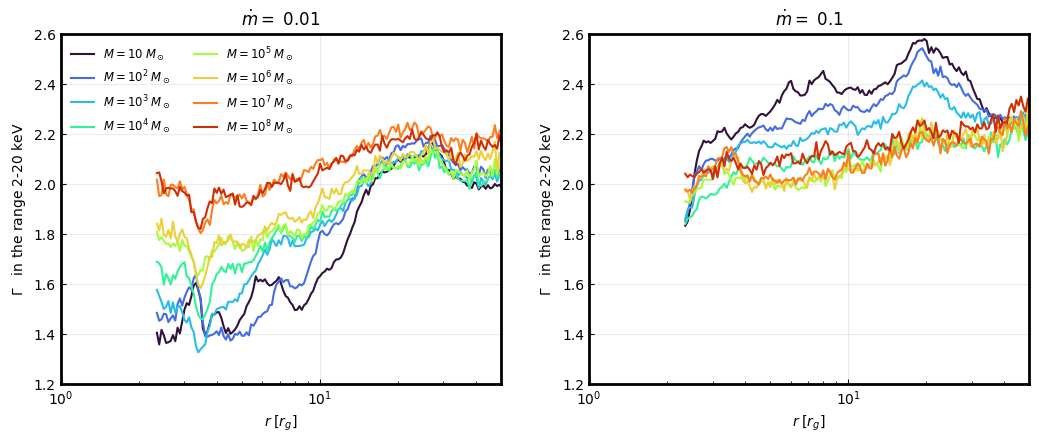}
    \caption{Azimuthally averaged photon index ($\Gamma$) of the flux incident on the disk measured at the photosphere in the energy range 2-20 keV. The left panel shows the $\mdot=0.01$ case and the right panel the $\mdot=0.1$ case while colors denote black hole masses. }
    \label{fig:Gamma_r}
\end{figure*}

\begin{figure*}
\includegraphics[width=\linewidth]{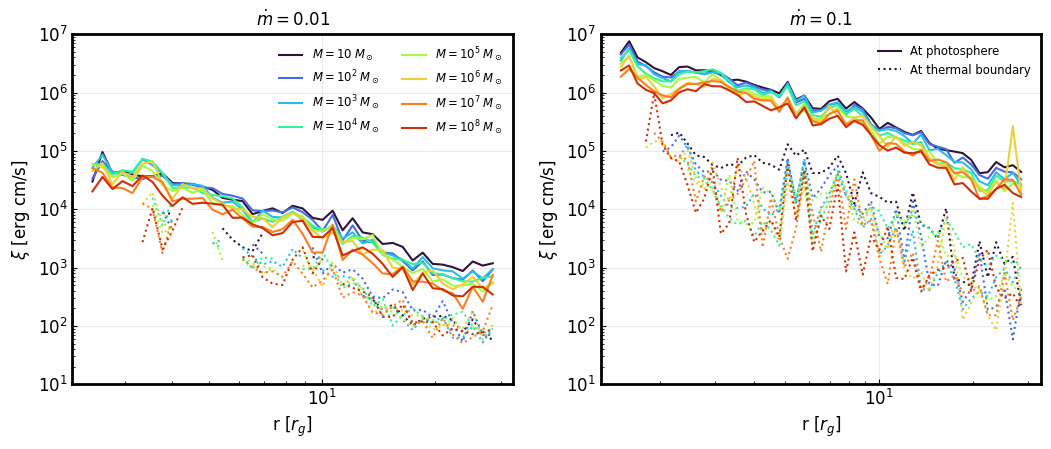}
    \caption{Azimuthally averaged ionization parameter ($\xi$) measured at the photosphere (solid line) and at the thermal boundary (dotted line). The left panel shows the $\mdot=0.01$ case and the right panel the $\mdot=0.1$ case while colors denoted black hole masses. }
    \label{fig:xi_r}
\end{figure*}

\begin{figure}
\includegraphics[width=\linewidth]{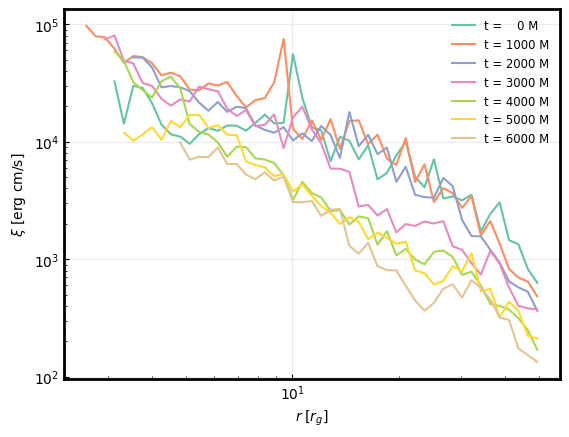}
    \caption{Same as Fig. \ref{fig:xi_r} for the models from \citet{Liu+2025}, where the time shown in the legend is the time after the inverse Compton cooling function is turned on \citep{Kinch_2020}.  }
    \label{fig:xi_paper_0}
\end{figure}

As described above, \texttt{PTransX} takes into account all the radiation processes relevant to X-ray reflection of an accretion disk, but to compute the reflected spectrum, it requires an external flux to serve as the boundary condition at the photosphere, providing a source of X-rays to be reflected. This flux is given by \texttt{Pandurata}, which self-consistently computes the radiation field outside the disk photosphere. The flux at the photosphere was presented in Figs. 8 and 9 of \cite{Nagele2026arXiv260103349N}, but we describe it again here. In the $\mdot=0.01$ case, the flux incident on the photosphere is well described by a broken power law with the break located at $\nu_{\rm break} = 100$ keV. Below the break, the power law is usually hard ($\Gamma<2$; here the spectrum is described as $F_\nu \propto \nu^{1-\Gamma}$) while above the break it is always soft ($\Gamma>2$).  The power law below the break becomes softer with increasing mass and increasing radius. For the $M=10^8\;\msun$ model, the power law has $\Gamma>2$ outside of $r=15\;r_g$. The slope above the break does not change with radius.

In the $\mdot=0.1$ case, the incident flux is qualitatively similar, but the power law has a low energy cutoff at about $30$ keV and is softer. Fig. \ref{fig:Gamma_r} shows the photon index as a function of radius measured in a narrower energy range relevant to Fe K$\alpha$ transitions, 2-20 keV (note that we do not consider an even narrower range of energies because of pollution of the power law slope from Fe K$\alpha$ photons incident on the disk). In this figure, we first measure $\Gamma$ at each location on the disk and then average in azimuth and hemisphere. $\Gamma$ increases at larger radii where the flux is generally softer, and the slope of $\Gamma(r)$ is greater for the lower mass models, particularly out to $r=20\;r_g$.

The degree to which the disk is ionized by this incident radiation can be estimated using the ionization parameter:
\begin{equation}
    \xi = \frac{4 \pi \int_{7.1 \;\rm{keV}}^\infty F_{\rm{up\;\&\;down}} d\nu}{n_e},
    \label{eq:xi}
\end{equation}
which can be thought of as the ratio of the number of ionizing photons to the number of ions times the mean photon energy. Note that $n_e$ is the electron number density, usually measured at the disk photosphere or the midplane, although in this paper we will focus on the scattering photosphere and the effecting photosphere. In the majority of the literature (including \citet{Nagele2026arXiv260103349N}), only the downward flux is included in the integrand. Here, however, we tally both the upward and downward flux, as the upward X-ray flux can be comparable to the downward
if there is significant reflection in the disk, or if the disk is only mildly optically thick ($\tau\sim1-4$), X-rays from the opposite photosphere can also contribute to ionizing events.

Fig.~\ref{fig:xi_r} shows the radial dependence of the azimuthally averaged ionization parameter at both the disk's photosphere and its thermal boundary.
The most important takeaway from this figure is that $\xi$ falls quite steeply with radius,
declining roughly $\propto r^{-2}$. 
We emphasize this point because $\xi$ is often assumed to be a constant in phenomenological spectrum-fitting. To further underline the robustness of this gradient in $\xi$,
we have computed $\xi(r)$ for seven different times within the ($\mdot = 0.01$,$M_{\rm BH} = 10 M_\odot$) simulation whose results were first reported in \citet{Liu+2025}.
Although the luminosity varied by a factor $\sim 3$ across this timespan,
Fig.~\ref{fig:xi_paper_0}, clearly shows that the radial trend shown in Fig.~\ref{fig:xi_r} 
persists virtually unchanged.  Only the overall scale of $\xi$ varies.
The other noteworthy feature of Fig. \ref{fig:xi_r} is that $\xi$ at the thermal core is generally $\sim 0.1 \xi$ on the photosphere at the same radius.

The second panel of Fig. \ref{fig:xi_r} shows the ionization parameter for the $\mdot=0.1$ case from \cite{Nagele2026arXiv260103349N}. The ionization parameter at the photosphere is well above 5000, which is a rough upper limit for where iron lines can be produced \citep{Reynolds2003PhR...377..389R}. Indeed, we showed in our previous paper that the $\mdot=0.1$ models did not have emission lines present in their spectra (especially at low mass, see Fig. 7). Observationally, the equivalent width of the Fe~K$\alpha$ line is seen to be about three times \textit{greater} in the soft state relative to the hard state \citep{Steiner2016ApJ...829L..22S}. This difference, however, is most readily attributed to 
the higher continuum flux near the line
in the hard state. 
There is no observational evidence for an intrinsic difference in the line flux 
of different spectral states. The fact that we see quite a large difference between emission line equivalent widths in the two accretion rate cases may point to missing physics in the GRMHD simulation (e.g. radiation pressure). 

There is, however, another potential systematic error in these equivalent widths.  They assume a line emission region confined within $30\;r_g$ of the black hole. This is because the underlying \texttt{HARM3D} simulations have achieved inflow equilibrium only out to $30\;r_g$. We run \texttt{PTransX} only in regions where inflow equilibrium has been achieved because in regions beyond this radius, the vertical density structure may still be transient and dependent on our initial data. Extrapolating the line surface brightness profile for $\mdot = 0.01$ suggests that the total line luminosity is reached near or a short distance beyond this radius; the same exercise for $\mdot = 0.1$ indicates that most of the line luminosity is made at larger radii.  Although including larger radii could change the equivalent widths, it may have a weaker effect on the line profile shapes because Doppler boosting scales relatively slowly with radius ($\propto r^{-1/2}$) and Compton broadening should vary much more slowly.  For all these reasons, in the rest of this paper, we will focus on the $\mdot = 0.01$ case.

\subsubsection{Ionization balance} \label{results:disk_flux_atomic}

\begin{figure}
\includegraphics[width=\linewidth]{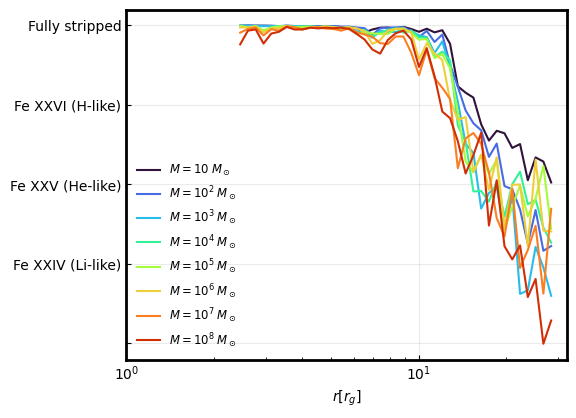}
    \caption{Average Fe ionization state as a function of slab radius. The average has been performed over mass density. Colors denote different masses and all models are from the $\mdot=0.01$ case.}
    \label{fig:Fe_ion_abundance_r}
\end{figure}

\begin{figure*}
\includegraphics[width=\linewidth]{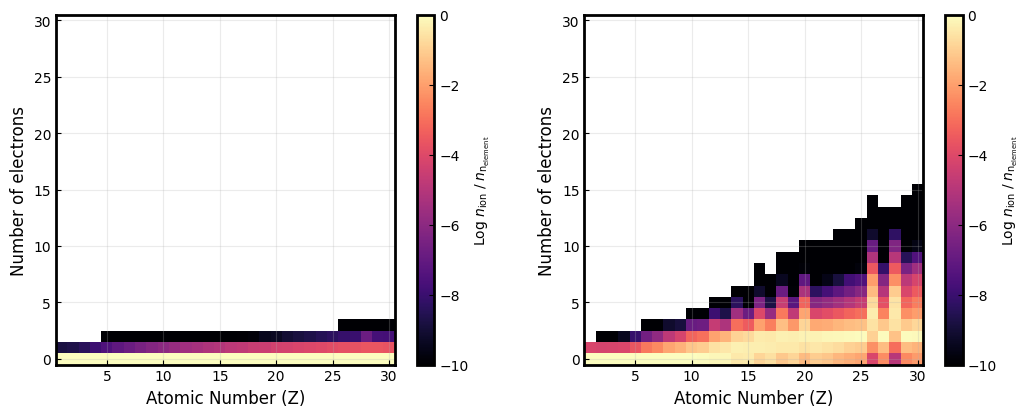}
    \caption{Two extremal examples of ion densities determined by \texttt{XSTAR} during the \texttt{PTransX} calculation. The horizontal axis is atomic number (e.g., Z(Fe) = 26, Z(Ni) = 28) and the vertical axis is the remaining number of electrons (0 electrons means fully-stripped, 1 electron H-like, etc). The colorbar shows $\log n_{\rm ion}/n_{\rm element}$, so that each column sums to 1. The left panel shows the ionization state of a cell near the surface of a highly ionized slab, located at $r=3.13 \;r_g$, whereas the right panel shows the ionization state of a cell with low ionization parameter ($\xi\sim100$, Fig. \ref{fig:xi_r}); this cell is located near the thermal boundary in a slab at $r=23.7\;r_g$. Both slabs are taken from the $M=10\;\msun$, $\mdot=0.01$ case. Other slabs from this model fall between these two extreme cases, so there are no locations with $r < 30 r_g$ in which the ionization density is dominated by ions with more than a few electrons. }
    \label{fig:ionization_balance_examples}
\end{figure*}

In each cell of the \texttt{PTransX} calculation at each temperature update, we use the photoionization code \texttt{XSTAR} \citep{Kallman2001ApJS..133..221K} to compute the ionization balance in that cell as a function of incident flux, density, and temperature. In this section, we describe the ionization fractions derived from the \texttt{XSTAR} computation. We emphasize that this study is the first to perform an ionization balance calculation using an incident flux that varies as a function of 3D position and time,
and whose spectrum is derived from specific radiation processes and a radiation transfer solution, rather than
an assumed power law. 
In fact, we have described above how
the spectrum illuminating the disk
is not well described by a single power law. There are some cases where even a broken power-law is not a good approximation, as there is curvature in the lower energy power-law like component (e.g. Fig. 9 of \citealt{Nagele2026arXiv260103349N}). 

In addition, and crucially, both the bolometric flux and the spectrum vary strongly with radius.

Fig.~\ref{fig:Fe_ion_abundance_r} shows the radial dependence of the mass-weighted average ionization
state of iron as a function of radius for each black hole mass in the $\mdot=0.01$ case. Inside $r=10\;r_g$, iron is almost always fully stripped. This result follows naturally from the left panel of Fig.~\ref{fig:xi_r}, which shows that inside this radius, $\xi>5000$ at the photosphere.
From $10 - 15 r_g$, the H-like state generally dominates.
Outside of this radius, the ionization states for different black hole masses diverge, with higher masses experiencing softer incident fluxes, which leave more of the Fe in He-like and Li-like states. Despite the ionization parameter dipping below 100 at the base of some slabs, the average composition is always highly ionized. 

This is also the case for elements other than iron. Fig.~\ref{fig:ionization_balance_examples} shows two extremal examples of the ionization state fractions for all the elements considered here.
The slab in the left-hand panel is located at $\simeq 3r_g$ and is very optically thin: $\tau_T < 1$.  Thus, it has copious X-ray illumination from both directions. These X-rays photoionize all the elements, so that even the heavier elements (Fe, Ni, etc.)
are more than 99$\%$ fully stripped. 
The right panel of Fig.~\ref{fig:ionization_balance_examples} shows a a cell deep within a slab at $\simeq 14r_g$, where $\xi$ reaches a minimum for this model.  As a result of low $\xi$, the stripped fraction is relatively low, although all of the elements are still highly ionized.
Below $Z=10$, the elements are still $>99\%$ stripped,
but even for much higher $Z$, the ionization balance shifts by only $\simeq 2 - 4$ electrons.
In fact, nowhere within $r \simeq 30\; r_g$ is Fe so weakly-ionized that the K$\alpha$ line appears at 6.4~keV.

\subsubsection{Atomic absorption edges} \label{results:absorption}

\begin{figure*}
\centering
\includegraphics[width=0.95\linewidth]{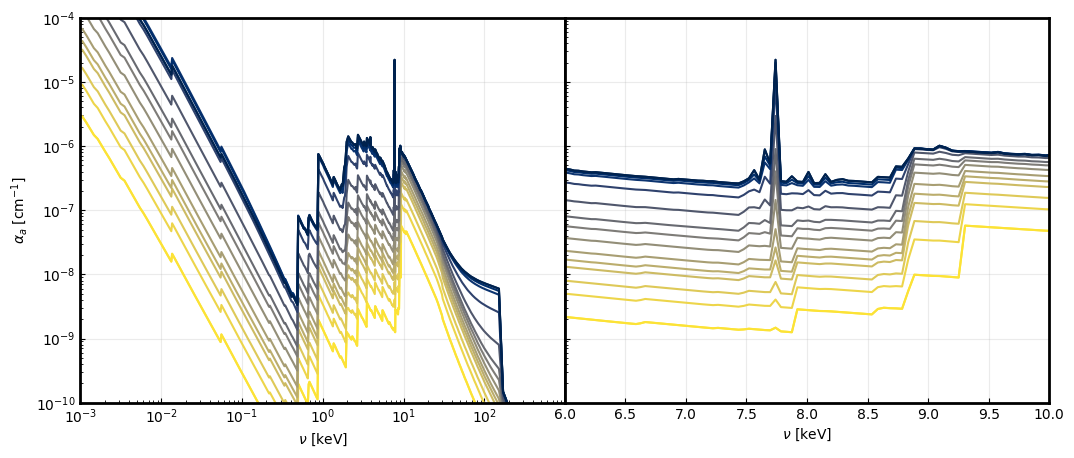}
    \caption{Spectral absorption coefficient $\alpha_a$ for a slab located at $r=24\;r_g$ in the $M=10\;\msun$ model. $\alpha_a$ includes both atomic transitions and bremsstrahlung. Each curve shows the coefficient for a different cell of the slab. The curve color corresponds to optical depth with brighter curves closer to the photosphere ($\tau=1$) and darker curves closer to the thermalization photosphere ($\tau=3.3$). (Left panel) Broad-band view. (Right panel) Zoom-in view on the Fe~K region of the spectrum.
    }
    \label{fig:absorption}
\end{figure*}

Fig.~\ref{fig:absorption} shows the absorption coefficient in different cells of a single slab located at $r=24\;r_g$ in the $M=10\;\msun$ model. The brighter curves at the bottom are closer to the photosphere, while the darker curves are deeper within the slab near the thermalization surface. Atomic absorption edges dominate the spectrum between 0.6 and 10~keV, but below this energy range, bremsstrahlung is dominant. 

Fig.~\ref{fig:absorption} shows a single example to display the main atomic features creating absorptivity. 
The quantitative relationships of these features vary with
 radius and black hole mass (and hence, ionization state). In the innermost regions of the disk, the absorption coefficient is lower overall due to smaller populations of partially stripped ions (see left panel of Fig.~\ref{fig:ionization_balance_examples}), and there are fewer absorption edges, although the largest absorption edges in Fig.~\ref{fig:absorption} remain. As one travels outward in radius, more absorption edges emerge, but this is not a monotonic trend as ionization balance depends on many factors. As black hole mass increases, the atomic absorption features remain roughly constant, but free-free absorption falls off as the density decreases proportional to $1/M$. The feature at $\nu=7.72$ keV is the Fe XXIII K$\beta$ line \citep{Klapisch1977JOSA...67..148K}. This transition is the strongest absorption feature of any Fe ionization state below Fe XXV \citep{Kallman2004ApJS..155..675K}.

\subsubsection{Disk internal radiation} \label{results:disk_flux_internal}

\begin{figure*}
\centering
\includegraphics[width=0.95\linewidth]{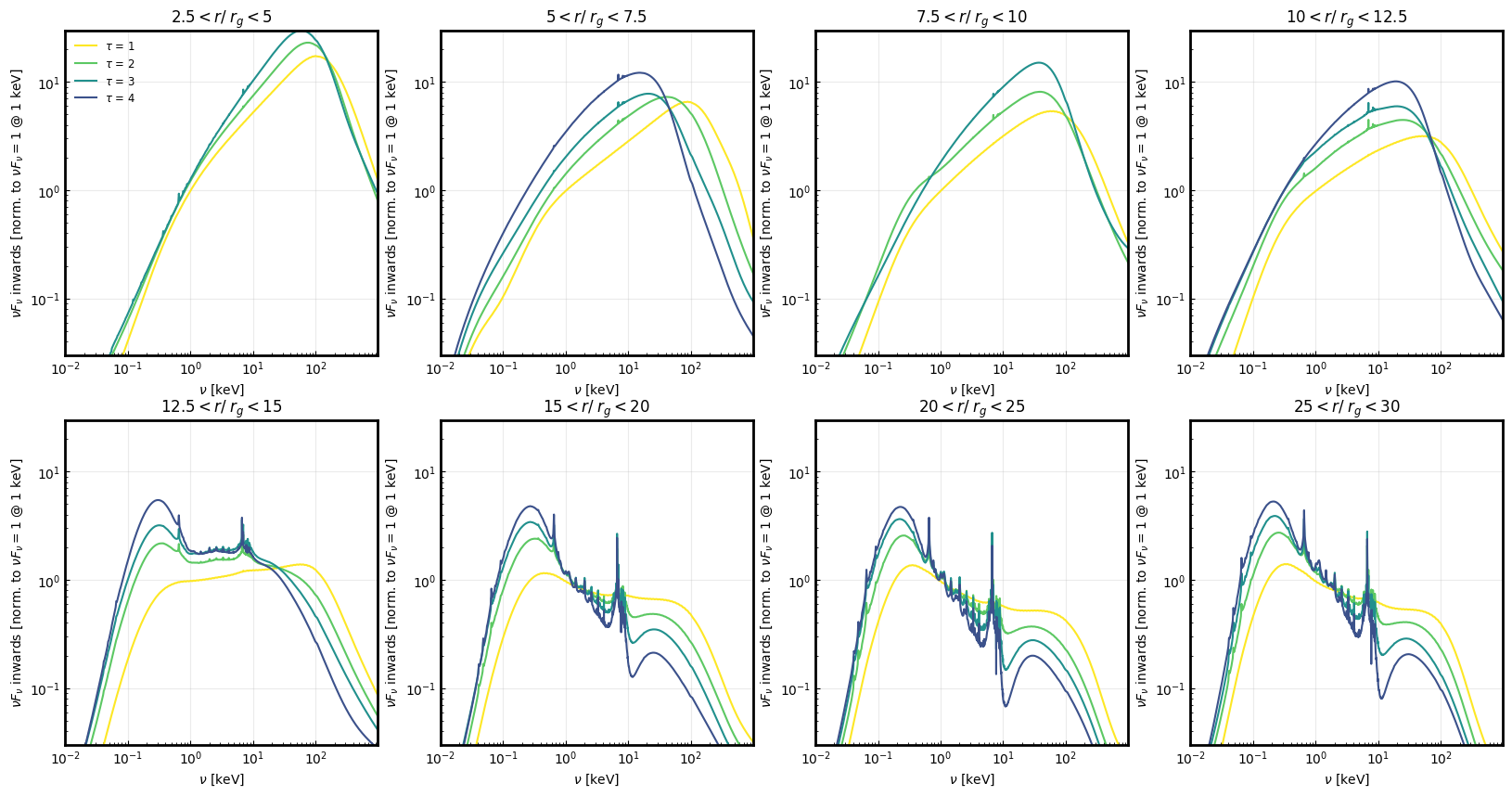}
    \caption{Inwards flux (traveling from the upper photosphere towards the lower photosphere/thermal core) at integer optical depths for the indicated radial bins in the $M=10\;\msun$, $\mdot = 0.01$ model. The flux at 1 keV at the photosphere ($\tau=1$) is normalized to unity.}
    \label{fig:flux_0.01_dn}
\end{figure*}

\begin{figure*}
\centering
\includegraphics[width=0.95\linewidth]{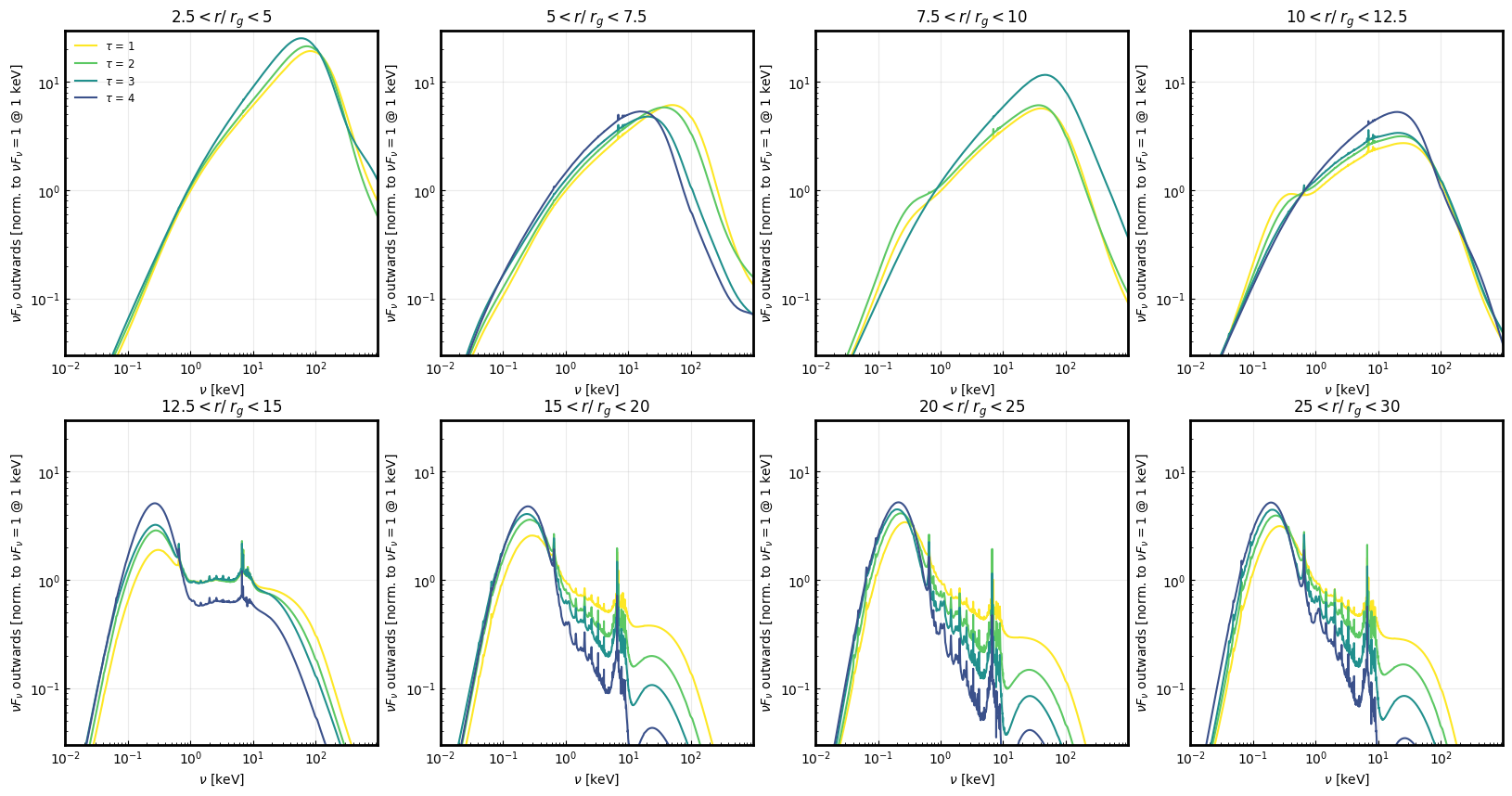}
    \caption{
    Same as Fig. \ref{fig:flux_0.01_dn} but for the outwards flux (traveling from the lower photosphere/thermal core towards the upper photosphere). The flux at 1 keV at the photosphere ($\tau=1$) is normalized to unity.
    }
    \label{fig:flux_0.01_up}
\end{figure*}

In this subsection, we describe the radiation flux inside \texttt{PTransX} slabs. Although this quantity is ancillary to the final spectrum, this exercise is useful for building an intuition of where emission lines are created and what determines their equivalent widths. Figs.~\ref{fig:flux_0.01_dn} and \ref{fig:flux_0.01_up} show spectral fluxes at integer optical depths averaged over all slabs in a particular radial bin (each panel shows a different bin) for the $M=10\;\msun$, $\mdot=0.01$ model. To avoid confusion, both figures are restricted to the northern hemisphere of the simulation. Fig. \ref{fig:flux_0.01_dn} shows the inwards flux and Fig. \ref{fig:flux_0.01_up} shows the outwards flux. In both cases, the yellow line labeled $\tau=1$ is the photosphere and lines with $\tau>1$ are inside the slab. All fluxes have been normalized so that $\nu F_\nu=1$ at $1$ keV for $\tau=1$. 

In Fig. \ref{fig:flux_0.01_dn}, the $\tau=1$ line is the incident flux onto the disk. As described above, this flux is roughly a power law below 100~keV, whose slope softens at larger radii. As we follow the averaged spectrum deeper into the slab, it loses flux at high energies and gains flux at low energies. This is due primarily to Compton scattering, but also to absorption and re-emission of high energy photons by ions. In the fifth panel ($12.5<r/r_g<15$), we begin to see a thermal peak at $\nu_{\rm peak}\sim0.2$ keV as some of the slabs become closer to thermal. This peak becomes more prominent at even larger radii. Fig.~\ref{fig:flux_0.01_up} shows the analogous outwards flux.  In both~Figs.~\ref{fig:flux_0.01_dn} and \ref{fig:flux_0.01_up}, the last three panels consist of almost entirely upper slabs which have thermal flux at the lower boundary, whereas the first two panels consist almost completely of whole slabs which have \texttt{Pandurata} flux at the lower boundary. In the region $7.5<r/r_g<15$, there is a mix of the two slab types.

We now shift our attention to the H-like Fe~K$\alpha$ line at $\sim6.97$ keV in Fig. \ref{fig:flux_0.01_dn}, which shows that even in the first panel, there is an iron line at $\tau>1$ (although it is minuscule). In the subsequent panels, oxygen and nickel lines appear. The equivalent widths of all emission lines grow monotonically with radius until the final three panels, which have similar spectra. As discussed above, the innermost radii are too highly ionized to produce abundant emission lines, and the radial dependence of the ionization states (Figs. \ref{fig:xi_r},\ref{fig:Fe_ion_abundance_r}) naturally predicts the smoothly increasing equivalent widths at larger radii.

We note that the emission lines have larger widths deeper in the disk. As these emission lines propagate out of the disk, they become diluted by both Compton broadening (see Appendix Fig. \ref{fig:flux_xi_sorted_zoom}) and an increasing level of the continuum. The increase in the continuum is from photons that entered the disk from above, penetrated to optical depths below that of the relevant spectrum, and were backscattered.
Because the surface illumination is so large, this effect raises the continuum flux, thereby diluting the emission line.

\subsubsection{Disk surface brightness} \label{results:disk_flux_surface}

\begin{figure}
\includegraphics[width=\linewidth]{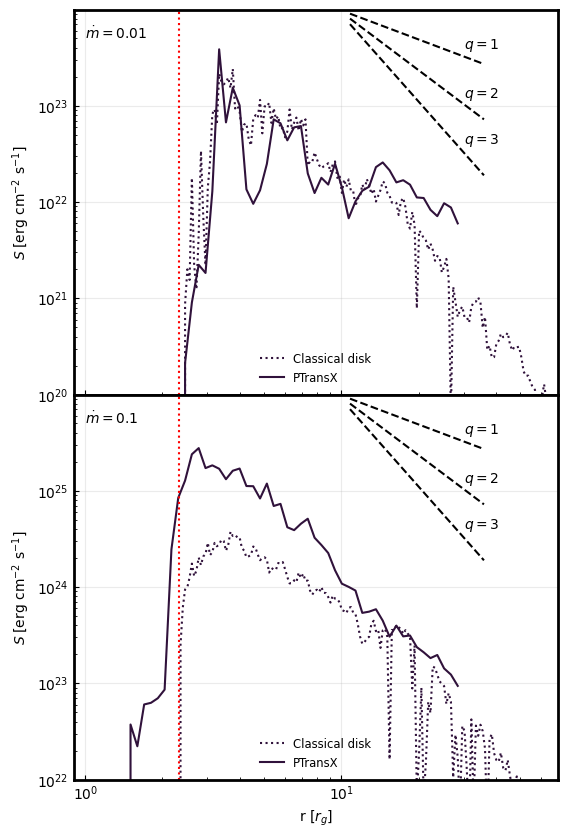}
    \caption{Surface brightness of the \texttt{HARM3D} disk according to classical accretion disk theory with the Shakura-Sunyaev reduction factor (dotted line) and \texttt{PTransX} (solid line). This figure shows $M=10\;\msun$ for the $\mdot=0.01$ case (upper panel) and the $\mdot=0.1$ case (lower panel). The vertical dotted line is the ISCO, while the dashed lines show integer emissivity indices. }
    \label{fig:surface10}
\end{figure}

\begin{figure}
\includegraphics[width=\linewidth]{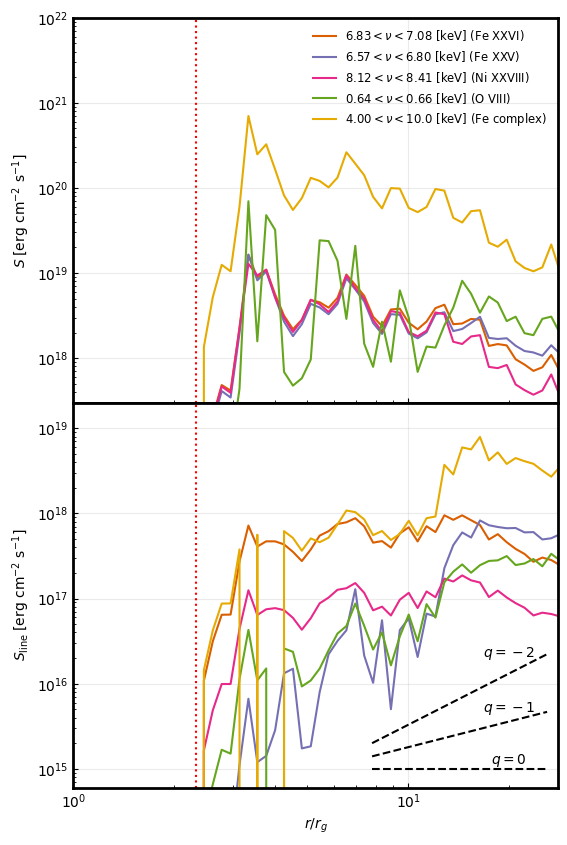}
    \caption{Surface brightness of several emission lines for the $M=10\;\msun$, $\mdot=0.01$ model. The upper panel shows the surface brightness in energy bands defined in the legend. The lower panel shows the continuum subtracted surface brightness in each band. Although the total surface brightness decreases with radius, the line surface brightness increases to a peak at around $r\sim 12-15\;r_g$. The vertical dotted line shows the ISCO. Note that the sum of the fluxes in the narrow iron and nickel lines is less than the flux in the 4-10 keV band. This is because the energy bands for the individual lines are too narrow 
    to include the Compton broadened component, but this 
    component 
    is contained within the 4-10 keV band (cf. Fig. \ref{fig:Ldisk_observed}). 
    }
    \label{fig:surface_lines}
\end{figure}

\begin{figure}
\includegraphics[width=\linewidth]{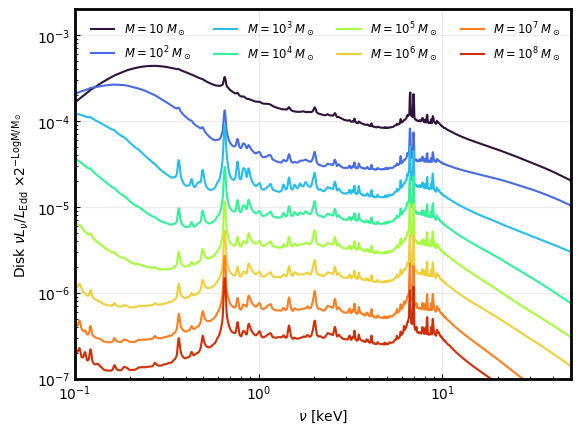}
    \caption{Disk luminosity (pre-\texttt{Pandurata}) as a function of mass in the $\mdot=0.01$ case. The curves have been separated by a factor of $2^{-\rm{Log} M/\msun}$ for legibility. }
    \label{fig:Ldisk_M}
\end{figure}

The surface brightness (i.e. the integrated flux [erg/s/cm$^2$]) is a ubiquitous diagnostic for understanding Fe~K$\alpha$ emission lines. In phenomenological modeling of observed spectra, it is often taken to scale as a power of the radius $S \propto r^{-q}$,
but set to zero for radii inside the ISCO.
When this method is applied to relativistically broad Fe~K$\alpha$ lines, $q$ is often found to be large.  In this subsection, we will describe the radial-dependence of the K$\alpha$ surface brightness predicted by physical simulations.

Before showing the Fe~K$\alpha$ surface brightness profile, we will first describe the bolometric surface brightness. Fig.~\ref{fig:surface10} shows the azimuthally averaged surface brightness at the photosphere as a function of radius. The two panels correspond to the two accretion rates for the $M=10\;\msun$ models (other masses have similar surface brightness profiles, due to energy conservation). We show the $\mdot=0.1$ case for comparison, but since it has too large of an ionization parameter to produce emission lines, we do not discuss it further and focus on the $\mdot=0.01$ case. The dotted line is the surface brightness 
calculated from the classical theory of time-steady accretion disks assuming the local accretion rate with the Newtonian reduction factor \citep{Shakura1973A&A....24..337S}, modified so that only the dissipation which occurs inside the $\tau=1$ scattering photosphere 
contributes to the flux \citep{Nagele2026arXiv260103349N}.
The solid line is the result from \texttt{PTransX}. The dashed lines show different emissivity indices (q = 1,2,3). In the $\mdot=0.01$ case, $q\leq 1$.

The bolometric surface brightness, however, can't be translated into an emission line surface brightness because it contains no information on the luminosity's energy-dependence. Fig.~\ref{fig:surface_lines} shows the surface brightness in narrow energy bands centered on the rest-frame energies of several emission lines including Fe~K$\alpha$; their radial-dependence is roughly $\propto r^{-1}$ for $r \gtrsim 3r_g$.
However, most of this luminosity in each of these energy bands is actually continuum.
We subtract off the continuum to arrive at the lower panel of Fig.~\ref{fig:surface_lines}, which shows the \textit{emission line surface brightness}, $S_{\rm line}$. Several notable emission lines are shown, including H-like and He-like Fe K$\alpha$ (6.7 and 6.97 keV, respectively), as well as oxygen and nickel emission lines. Finally, the gold line shows the entirety of the Fe K$\alpha$ region, which will include the aforementioned iron and nickel lines, as well as small contributions from Cr, Mn and others. All of the emission lines in this energy range may contribute to a relativistically broadened line, as we will discuss.  

The surface brightness of all lines have positive slopes ($S_{\rm line} \propto r^{0.5}-r^{2}$).
Integrating the emission line surface brightness over the area of the photosphere, our findings indicate that the dominant source for all the lines, including Fe~K$\alpha$, is found at radii at least $\sim 10 - 30 r_g$.

Fig. \ref{fig:Ldisk_M} shows the total disk luminosity 
in Eddington units for each mass in the $\mdot=0.01$ case. The oxygen, iron and nickel lines discussed above can be seen for each model. 
The spectra in this figure are the summation of the seed photon spectra injected into the corona.

As previously emphasized for both the continuum \citep{Krolik2002ApJ...573..754K,Schnittman_2013} and the Fe~K$\alpha$ line \citep{Kinch_2016,Kinch_2019,Kinch_2021}, the radial profile of the surface brightness generally has an inner edge somewhere in the vicinity of the ISCO, but coincides closely with the ISCO only in special cases.  For example, we find that the continuum surface brightness follows closely the classical prediction for $\mdot=0.01$, but differs substantially for $\mdot=0.1$ (Fig.~\ref{fig:surface10}).  For the K$\alpha$ line with $\mdot=0.01$, the inner edge for FeXXV lies at $\simeq 15r_g$, while for FeXXVI it is at $\simeq 4r_g$; both of these are well outside the ISCO.

\subsection{Observed flux} \label{results:obs_flux}

Up until now, we have focused on
spectra in the fluid frame. In this 
subsection we will present the spectra seen by distant observers, 
including all relativistic effects.
First, we will describe the observed shape of a single emission line. Next, we will generalize that description to the total disk flux. Finally, we will incorporate Compton scattering and disk reflection of photons in the corona, and describe
emission lines as they are observed, i.e., the features by which they contrast with the adjacent continuum.
In this subsection we will focus on the broad Fe~K$\alpha$ line with figures displaying the energy range 4 to 10 keV. Similar analyses could be carried out for lower energy lines (such as oxygen lines, cf. Fig.~\ref{fig:surface_lines}), but such features are observed 
much less frequently.

\subsubsection{The observed spectrum from a single emission line} \label{results:obs_flux_line}

\begin{figure}
\includegraphics[width=\linewidth]{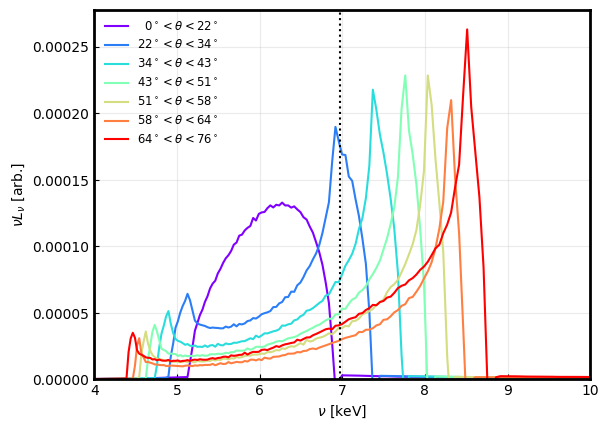}
    \caption{Observed spectrum at different polar viewing angles  of the \texttt{Pandurata} spectrum coming from the photosphere of a single column (at $r =10\; r_g$) with a single injected line (black dotted line at $\nu_i=6.97$ keV).
    Angles in the legend are in degrees.
    The observed spectra show the characteristic asymmetric double horned pattern seen off-axis due to
    varying combinations of boosting and gravitational redshift.
    }
    \label{fig:double_horned}
\end{figure}

\begin{figure}
\includegraphics[width=\linewidth]{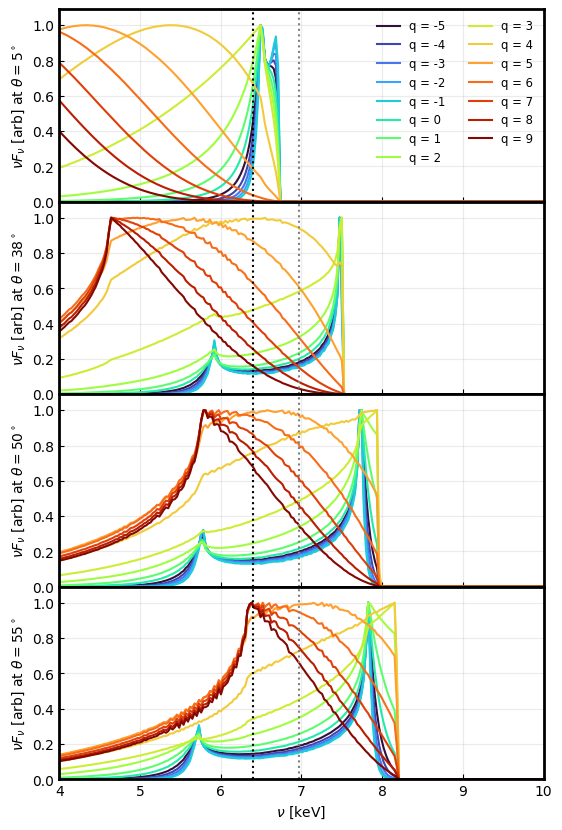}
    \caption{Normalized line profiles 
    for a single emission line at 6.97 keV (second dotted vertical line) coming from a disk 
    whose line surface brightness extends from
    $r=r_{\rm ISCO}=2.7\;r_g$ to $r=30\;r_g$
    and is $\propto r^{-q}$. The panels correspond to viewing angles 
    $\theta=5^\circ$, $\theta=38^\circ$, 
    $\theta=50^\circ$, $\theta=55^\circ$.
    At these angles, the line is seen to peak at 6.4~keV (first dotted line, the energy at which observed lines often peak) when $q=0-3$, $q=4$, $q=5$, and $q=7-9$, respectively.
    This figure illustrates the strong dependence of the flux peak on viewing angle for a given spin parameter and emissivity profile. }
    \label{fig:Lineart_single}
\end{figure}

To illustrate the several effects combining to produce the shapes of observed emission line profiles, we begin with the simplest case, effects due to photon propagation along geodesics. We will then add complexity piece by piece. In
Fig.~\ref{fig:double_horned}, we show how the profile of
an emission line injected at a single energy ($\nu_i=6.97$ keV) from a single point on the \texttt{HARM3D} photosphere ($r=10\;r_g$) 
varies with
polar viewing angle. For this figure, we have turned off Compton scattering in the corona and reflections of photon packets off the disk (any packet impacting the disk simply halts
and is not observed). The spectra in Fig.~\ref{fig:double_horned} exhibit the well known double horned pattern \citep{Reynolds2003PhR...377..389R}. This pattern arises from the combination of gravitational redshift, which shifts the observed line to lower energy, Doppler boosting, which produces the two peaks of the line at inclined viewing angles (referred to as the red and blue horns), and causes the blue horn to have more power than the red horn.

Next, we generalize from emission at a single radius to emission over the entire disk. 
For a pedagogical example, we follow the phenomenological practice already described and assume
the line surface brightness is zero inside the ISCO and
outside that radius is $\propto r^{-q}$.
For ease of comparison with our post-processing procedure, we will assume zero surface brightness for $r > 30r_g$.  A larger cut-off radius would create significantly different profiles when $q \lesssim 3$.
The line profiles corresponding to surface brightness profiles of this form are shown at selected viewing angles in Fig.~\ref{fig:Lineart_single} for $-5 \leq q \leq 9$, as computed with
the open source code \texttt{LineAART} \citep{Gates2025PhRvD.111l4004G}.
We require the disk to be optically thick so as to block photon trajectories passing multiple times around the black hole.
The panels show several different polar viewing angles, 
for each of which certain values of $q$ have line peaks close to 6.4~keV, the energy of the K$\alpha$ line in neutral Fe.

\subsubsection{
Setting the line against the rest of the disk spectrum}
\label{results:obs_flux_disk}

\begin{figure}
\includegraphics[width=\linewidth]{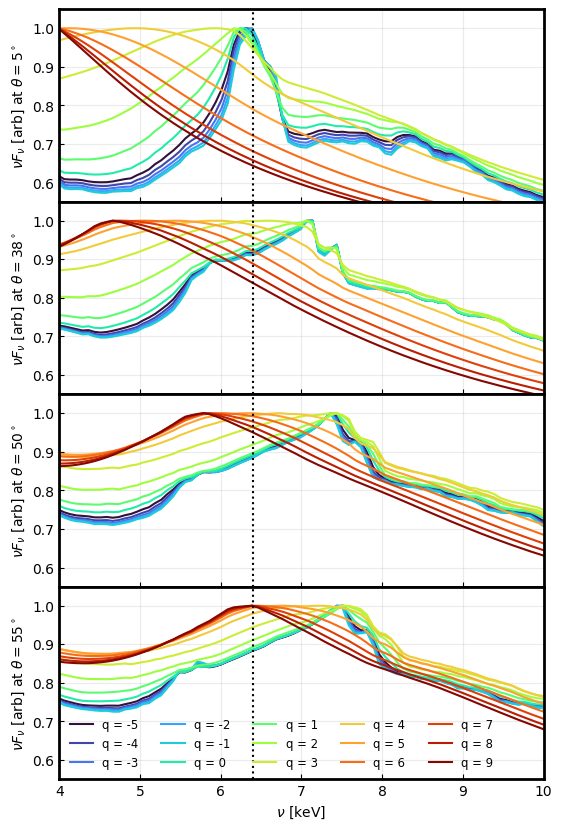}
    \caption{Same as Fig. \ref{fig:Lineart_single}, but instead of injecting a single emission line at $6.97$ keV, we inject the spectral disk luminosity of the $M=10\;\msun$, $\mdot=0.01$ model, which can be seen in Fig. \ref{fig:Ldisk_M}. This is not equivalent to the final \texttt{Pandurata} result because the surface flux has a variable spectrum (see, e.g. Fig. \ref{fig:surface_lines}).}
    \label{fig:Lineart_total}
\end{figure}

\begin{figure}
\includegraphics[width=\linewidth]{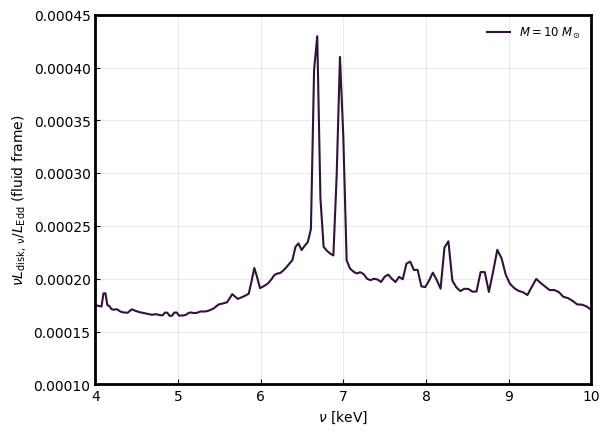}
    \caption{Disk luminosity in the fluid frame (pre-\texttt{Pandurata}, dark line, as in Fig. \ref{fig:Ldisk_M}) for the $M=10\;\msun$, $\mdot=0.01$ model. }
    \label{fig:Ldisk_observed}
\end{figure}

\begin{figure}
\includegraphics[width=\linewidth]{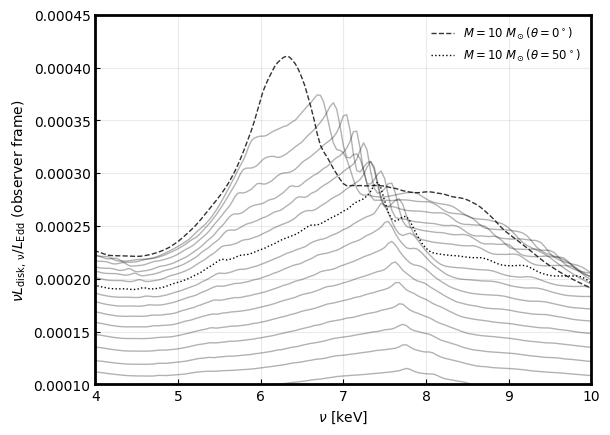}
    \caption{Disk luminosity in the observer frame for the $M=10\;\msun$, $\mdot=0.01$ model. The gray curves show individual polar viewing angles from the northern hemisphere, with the viewing angles separated by intervals of constant cos($\theta$). The dashed and dotted lines show two particular polar viewing angles: $\theta=0^\circ, 50^\circ$. The individual viewing angles have been scaled as if they were emitting over the entire sphere. 
    }
    \label{fig:Ldisk_sightlines}
\end{figure}

In Fig.~\ref{fig:Lineart_total}, we show the spectral shape
when the entire spectrum of seed photons is endowed with a surface brightness $S\propto r^{-q}$. This exercise is similar to the one in Fig. \ref{fig:Lineart_single}, but instead of injecting a single emission line at 6.97 keV, we inject the disk luminosity from Fig. \ref{fig:Ldisk_M}.
Specifically, we use the sum of local fluid-frame seed photon 
luminosities for $M=10\;\msun$.
We stress that for two reasons Fig.~\ref{fig:Lineart_total} does {\it not} show what a real observer would see:
it assumes a power-law radial profile for the disk surface brightness; and it does not allow for spatial variation in the shape of the seed photon spectrum.
Nevertheless, the line profiles in this figure should roughly resemble the real line profiles presented in the next section. 

At a nearly face-on viewing angle ($\theta=5^\circ$),
the line profiles for $q \leq 2$ all produce a broad feature peaking at slightly less than 6.4~keV and
extending down to $\simeq 4 - 5$~keV on the red wing and up to $\simeq 6.8-10$~keV on the blue wing.
The profile within $\pm 0.5$~keV of the peak is slightly asymmetric,
as the He-like Fe~K$\alpha$ line has slightly greater luminosity than the H-like K$\alpha$ line.
In contrast, the inclined viewing angles for this range of $q$ have two distinct peaks, both located between $7$ and $8$~keV, corresponding to the blue horns of the He- and H-like emission lines (cf. Fig.~\ref{fig:Lineart_single}). 
For higher values of $q$, i.e., surface brightness profiles declining sharply with increasing radius, viewers at $\theta \lesssim 50^\circ$ always see a peak at well below 6.4~keV, whereas more inclined viewers will observe a peak in the range $6.4-7.5$ keV. This single peak corresponds to the  blue horns of the two Fe K$\alpha$ emission lines, which are blended by the extreme gravitational redshift. There also exists a corresponding red horn at much lower energy (not shown in Fig. \ref{fig:Lineart_total}).

\subsubsection{Emission lines with full physics}
\label{sec:real_lines}

\begin{figure}
\includegraphics[width=\linewidth]{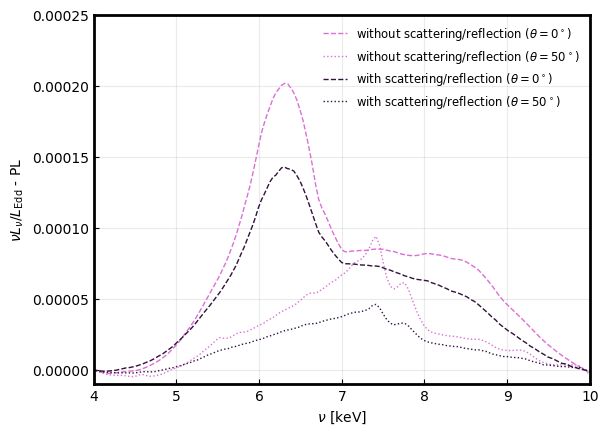}
    \caption{Residual line profiles for \texttt{Pandurata} spectra with scattering and reflection turned off (pink lines, cf. dashed and dotted lines in Fig. \ref{fig:Ldisk_sightlines}) and turned on (black lines). The dashed lines are for a face-on view ($\theta=0^\circ$) while the dotted lines are for an inclined view ($\theta=50^\circ$). }
    \label{fig:line_with_scattering}
\end{figure}

As a penultimate step toward the full physics of line-formation, we show the disk luminosity seen by both local and distant observers. In Fig.~\ref{fig:Ldisk_observed} we show a zoomed in version of Fig. \ref{fig:Ldisk_M}, focusing on the Fe K$\alpha$ emission region. We will show this energy range throughout this section while discussing the emission line. There are three features of note in this figure. First, there are two narrow iron emission lines, corresponding to He-like and H-like iron at 6.7 and 6.97 keV, respectively. Although we show a single composite luminosity here, note that the 6.7 keV emission line is generally produced at larger radii than the 6.97 keV emission line (Fig. \ref{fig:surface_lines}). The second feature is a Compton broadened iron line centered roughly at 6.7 keV or slightly lower energy. This feature is present because the iron emissivity at 6.4-6.7 keV increases deeper into the slab, where Fe becomes less highly ionized. The line photons generated in these regions undergo several electron scatterings as they travel outward, broadening the line by the time it reaches the photosphere. The third feature of interest is the smattering of iron and nickel lines between 7.5 and 9 keV, which we will discuss later. 

Fig.~\ref{fig:Ldisk_sightlines} shows the disk luminosity seen by a distant observer. We integrate \texttt{Pandurata} photon packets to infinity as if they were traveling through a transparent corona, which is the standard \texttt{Pandurata} procedure albeit with Compton scattering and disk reflection turned off. When photons impact the disk, they are terminated instead of being reflected. This figure is thus showing the luminosity of photons making their first exit of the disk. Fig.~\ref{fig:Ldisk_sightlines} shows spectra seen from polar angles ranging from $0^\circ$ to $90^\circ$ (grey curves). Two polar views ($\theta=0^\circ,50^\circ$) are denoted by the dashed and dotted grey curves. We have chosen $\theta=50$ as a prototypical inclined viewing angle because many phenomenological measurements of the inclination angle fall into $\theta \in [30^\circ,70^\circ]$. We will show these two viewing angles in many of the figures throughout the rest of the paper. 
For the iron emission line, the face-on view gives a luminous, broad, peak. As the viewing angle shifts toward more edge-on, this peak is broadens, shifts blueward, and rises less above the continuum\footnote{
The smoothness of the grey curves in this figure demonstrate that, even after being broken up into 41 bins, there is little Poisson noise in our spectra.  Our spectral predictions, which use $\sim 10^6$ photon packets, are therefore well-converged.}.

\subsubsection{The observed spectrum from the entire system} \label{results:obs_flux_system}

\begin{figure}
\includegraphics[width=\linewidth]{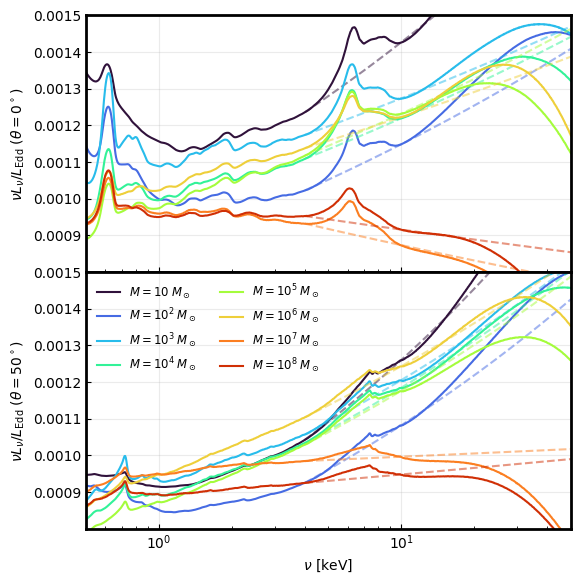}
    \caption{\texttt{Pandurata} spectra in the X-ray band for each mass in the $\mdot=0.01$ case. The upper panel shows the face-on view ($\theta=0$), and the lower panel  a viewing angle of $\theta=50$. The under-plotted dashed lines show the power-law fits between 4 and 10 keV (Fig. \ref{fig:line_with_scattering}). Some, but not all, models have Compton humps peaking around 20-40 keV. }
    \label{fig:Ltot}
\end{figure}

The last step toward complete line formation physics is to reintroduce Compton scattering and disk reflection.
Fig.~\ref{fig:line_with_scattering} compares the $\theta=0^\circ,50^\circ$ curves for the spectrum as seen through a transparent corona in Fig.~\ref{fig:Ldisk_sightlines} (pink curves) with the spectra at the corresponding viewing angles 
with full physics
(black curves).
To make this comparison, we plot an approximate version of what observers might call the line profile: the total flux as a function of energy minus an estimated continuum spectrum.
Our ``continuum" is found from a power law defined by the luminosity at $\nu=4$ keV and the luminosity at $\nu=10$ keV.
The dark curves, which include Compton scattering, have lower luminosity (relative to the continuum), as some Fe K$\alpha$ photons scatter on their way to the observer. The amount by which the line photons are attenuated
increases with greater inclination:
at $\theta=0^\circ$ the line flux falls by $25\%$ and by $50\%$ at at $\theta = 50^\circ$.

Fig. \ref{fig:Ltot} shows the total observed luminosity with all relevant physics over a wider range of energies. 
From top to bottom, this figure shows: an average over viewing angle; the face-on view; and a strongly-inclined view ($\theta = 50^\circ$).
The main effect of including Compton scattering and disk reflection is to change the continuum slope of the observed spectrum. Whereas in Fig.~\ref{fig:Ldisk_M}, the continuum declines (in $\nu L_\nu$ units) with increasing photon energy,
here the continuum rises ($M\leq10^6\;\msun$) or is roughly flat ($M\geq10^7\;\msun$). 
This contrast is, of course, the effect of inverse Compton scattering in the hot corona. This figure shows the remarkable fact that the emission line profile itself changes much less with black hole mass than the underlying continuum does.

\subsubsection{Emission line equivalent widths}
\label{sec:equivalent_widths}

\begin{figure}
\includegraphics[width=\linewidth]{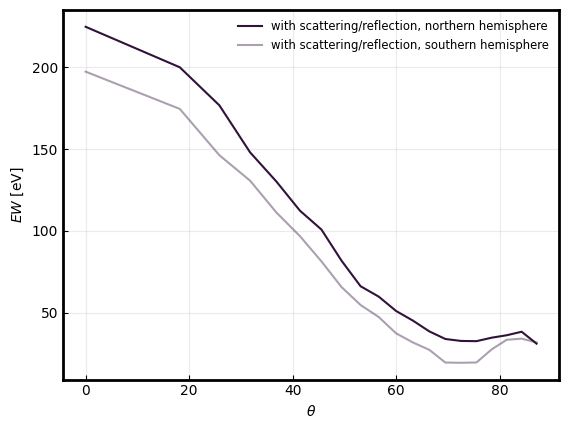}
    \caption{Equivalent width of the Fe K$\alpha$ line 
    as a function of polar viewing angle for the northern hemisphere (darker line) and the southern hemisphere (lighter line). Both curves correspond to the $M=10\;\msun$, $\mdot=0.01$ model. }
    \label{fig:equivalent_10}
\end{figure}

\begin{figure}
\includegraphics[width=\linewidth]{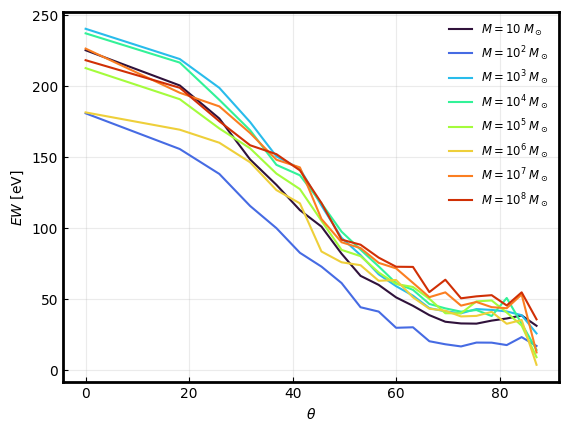}
    \caption{Same as Fig. \ref{fig:equivalent_10} but for all masses in the $\mdot=0.01$ case. }
    \label{fig:equivalent_all}
\end{figure}

Emission lines are often diagnosed by their equivalent widths:
\begin{equation}
  EW \equiv  \int d\nu\; \frac{F_{\rm line}-F_{\rm continuum}}{F_{\rm continuum}}.
\end{equation}
$F_{\rm line}-F_{\rm continuum}$ is the quantity shown in Fig. \ref{fig:line_with_scattering}, where $F_{\rm line}$ is the flux observed between two energies either side of the line centroid (cf. legend of Fig. \ref{fig:surface_lines}), and $F_{\rm continuum}$ is the power law flux fit to the observed flux at those two energies.
Using this definition, we compute the equivalent width as a function of polar viewing angle (Fig.~\ref{fig:equivalent_10}). The two lines show the widths measured by observers viewing the northern and southern hemispheres of the $M=10\;\msun$, $\mdot=0.01$ model. As expected, the width decreases monotonically with increasing angle \citep{Fabian2000PASP..112.1145F} from $\simeq 225$~eV for $\theta = 0^\circ$ down $\sim 25$~eV for $\theta \gtrsim 70^\circ$,
where the line has almost completely disappeared. 
At nearly edge on viewing angles, much of the flux originating from the disk does not have a clear line of sight to the observer. Future GRMHD simulations with more detailed physics at larger radii where the disk (and torus in the AGN case) play an important role in attenuating the flux from the inner disk will be required to truly understand the line profiles from these extreme viewing angles. 
This range of equivalent widths is well in line with observations (Sec. ~\ref{sec:discussion}), and we have not resorted to super-solar abundances to boost the equivalent widths \citep{Draghis2025ApJ...989..227D}.

Fig.~\ref{fig:equivalent_all} shows the same figure (northern hemisphere only) for all masses in the $\mdot=0.01$ case. As mentioned before, the iron line profiles are extremely similar for different masses, so the variation in the equivalent widths at fixed viewing angle (a range $\simeq 30 - 60$~eV) comes primarily from relatively small differences in the continuum level; as is plain in Fig.~\ref{fig:Ltot}, there is enough continuum curvature over scales of a few keV to make continuum placement delicate.

\section{Discussion} \label{sec:discussion}

\subsection{Profile categories of observed Fe K$\alpha$ lines}

There have been many observations of Fe~K$\alpha$ lines radiated from accreting black holes \citep{Tanaka1995Natur.375..659T,Nandra1997ApJ...477..602N,Miller2002ApJ...570L..69M,Boller2002MNRAS.329L...1B,Vaughan2004MNRAS.348.1415V,Fabian2009Natur.459..540F,Reis2011MNRAS.410.2497R,Brenneman2011ApJ...736..103B,Risaliti2013Natur.494..449R,Walton2014ApJ...788...76W,Walton2016ApJ...826...87W,Xu2018ApJ...852L..34X,Jia2022MNRAS.511.3125J,Draghis2023ApJ...946...19D,Draghis2023ApJ...947...39D,Zdziarski2025arXiv251103386Z}, and recently line profiles have been measured at high resolution with XRISM \citep{Draghis2025ApJ...995L..12D,Brenneman2025ApJ...995..200B,Chatterjee2025ApJ...992..210C,Li2026A&A...706A.255L,Liu2026arXiv260306883L}. These line profiles (including the recent XRISM observations) are usually singly peaked, and those lines that do have two peaks often have the secondary peak at higher energy than the main peak,
e.g., as in the X-ray binary V4641~Sgr \citep{Draghis2023ApJ...946...19D}, which suggests that peak multiplicity comes from multiple emission lines instead of Doppler boosting.

Some of these single-peaked lines lines are nearly symmetric in energy space \citep{Reis2011MNRAS.410.2497R,Xu2018ApJ...852L..34X,Jia2022MNRAS.511.3125J,Draghis2023ApJ...946...19D,Liu2026arXiv260306883L}. 
Our results suggest these lines are most likely from disks viewed close to face-on (Fig.~\ref{fig:Ltot}), a statement which is almost always in tension with the inclinations derived from phenomenological models. We discuss this tension further in Sec. \ref{sec:non_uniqueness}.

Others are asymmetric. In some, the peak is located at an energy higher than the line centroid \citep{Walton2014ApJ...788...76W,Draghis2023ApJ...946...19D,Brenneman2025ApJ...995..200B,Zdziarski2025arXiv251103386Z,Li2026A&A...706A.255L}.  When this is the case, our results agree with the standard interpretation of these observations \citep{Reynolds2021ARA&A..59..117R}: the lines originate from disks viewed at a more inclined angle.  These line profiles are similar to the $\theta=50^\circ$ curves from the previous section (dark curve in Fig. \ref{fig:line_with_scattering}, second panel in Fig. \ref{fig:Ltot}). 

Finally, some single-peak profiles have a peak at lower energy than the centroid \citep{Walton2016ApJ...826...87W,Draghis2023ApJ...946...19D,Draghis2023ApJ...947...39D,Draghis2025ApJ...995L..12D,Chatterjee2025ApJ...992..210C}.
None of the models studied here has features similar to this pattern. This is likely because we compute \texttt{PTransX} out to only $r=30\;r_g$ (Sec. \ref{results:disk_flux_incident}). The peak of the line profile often sits directly at 6.4 keV \citep[e.g.][]{Draghis2025ApJ...989..227D}, and is likely produced in the region $r>30\;r_g$. Currently, our rest frame spectra peak at 6.7 keV (Fig. \ref{fig:Ldisk_observed}), but if we were to add a rest frame component at 6.4 keV (originating at $r>30\;r_g$), even if that component were somewhat narrower it would shift the peak to the left of the line centroid.

\subsection{
Where in the disk is the Fe~K$\alpha$ line formed?}

The question of where in an accretion disk Fe K$\alpha$ photons can be produced is a long-standing one. Early on, \citet{Reynolds1997ApJ...488..109R} raised the possibility of Fe~K$\alpha$ photons emitted inside the ISCO, although this idea was 
contested
by \citet{Young1998MNRAS.300L..11Y}. \citet{Krolik2002ApJ...573..754K} pointed out that the \textit{reflection edge} (outside of which reflection features can be seen) could feasibly fall on either side of the ISCO, depending on the accretion rate.
Other MHD simulations \citep{Reynolds2008ApJ...675.1048R} indicated that---for a point source corona with parameters popular at that time---the reflection edge was located within 0.2 $r_g$ of the ISCO, but only if the density at the photosphere was artificially raised by a factor $\sim 50$ to bring $\xi$ below $\sim 5 \times 10^3$.  Their motivation for this alteration was to find a way to create K$\alpha$ lines of the observed amplitude.  As we will show explicitly in Sec.~\ref{sec:non_uniqueness}, detailed photoionization calculations provide another solution.
Many of these previous studies expected that the reflection edge should be close to the ISCO.   By contrast, in the examples treated in this work we find that it is located at $\approx 12\;r_g$ even though, for the relatively high spin modeled here, $r_{\rm ISCO} = 2.3 \;r_g$ (Fig. \ref{fig:surface_lines}).   

More recent studies have estimated the spatial variation scale of the ionization parameter using density variation scales from GRMHD simulations \citep{Shashank2022ApJ...938...53S,Mirzaev2024ApJ...976..229M,Shashank2025PhRvD.112l3030S}. They concluded that in order for there to be partially ionized atoms near the ISCO, the overall density scale had to be higher than was typically thought,but did not suggest any mechanism that would raise the density  to this level. 
Although our treatment of thermodynamics in the disk body can certainly be improved, the density found by our simulation results from solving the dynamical equations consistent with its thermodynamical assumptions. Moreover, as we have shown, it points out a different way to make broad Fe~K$\alpha$ profiles.

Where the K$\alpha$ emission arises is central to the shape of Fe K$\alpha$ profiles because the relativistic effects depend on the emitting location. This question can be rephrased in more physical language by posing it in terms of the radial-dependence of three quantities determining whether a particular emission line can be made: the ionization parameter, which is proportional to the incoming flux; the photon index, which encodes the slope of the incoming flux spectrum (cf. the ratio of Fe K-edge photons to H-ionizing photons); and the surface density ($\Sigma = \int_{\rm disk} \rho dz$), which is closely related to the $K$-edge optical depth. To see the importance of this last quantity, consider the limiting cases: as $\Sigma \to\infty$, then an iron line can always be emitted, as long as $\log \xi > 0$. As $\Sigma \to 0$, then an iron line cannot be emitted as there are no iron ions to do the emitting.

Regrettably, our calculations cannot offer a firm answer to the question of where most of the K$\alpha$ luminosity is generated because in many cases we find the emissivity remains large at $30r_g$, the outer bound of the region whose K$\alpha$ emission we compute.  This bound is placed by our best estimate of the radius within which the surface density is governed more by physical evolution than the initial condition we chose.  Because equilibration in this sense is achieved on a timescale roughly $\propto r^{3/2}$, The only remedy is \textit{much} longer duration simulations.

\subsection{Limitations of phenomenological modeling}

The rephrasing just described lies at the heart of phenomenological modeling of disk-surface ``reflection"
\citep{Ross2005MNRAS.358..211R,Garcia2010ApJ...718..695G,Garcia2014ApJ...782...76G,Huang2025arXiv251212728H}. In many applications of this method a strong assumption is made about the spatial-dependence of these quantities: that there isn't any, i.e., all three are the same for every location in the disk\footnote{To be more precise about the surface density, the actual assumption is that the Compton optical depth at every location is $\geq 10$, and the density does not vary vertically anywhere in the 10 optical depths below the photosphere.} (e.g. \citealt{Reynolds2021ARA&A..59..117R}, although see \citealt{Zhang2024MNRAS.532.3786Z} for a possible spatial-dependence of the spectral slope). In some other studies \citep[e.g.][]{Dauser2013MNRAS.430.1694D}, the ionization parameter does vary, but only according to simple unphysical models such as the lamppost corona which assumes all X-ray irradiation originates from a point source above the black hole. 
In several other studies \citep{Abdikamalov2021ApJ...923..175A,Abdikamalov2021PhRvD.103j3023A,Mall2022MNRAS.517.5721M}, the ionization parameter and constant disk density were allowed to vary freely as power laws, but the ranges of these parameters did not cover the relevant regions which we will discuss below (these studies were limited to $\xi_{\rm in} \leq 10^4$, $\alpha_n\geq 0$, cf. Table \ref{tab:slopes}).

Our work calls into question the validity of the simple assumptions described above.  In Figure~\ref{fig:Gamma_r}, we showed that the spectral slope of the flux incident on the disk generally becomes steeper (softer) with increasing radius.   In Figure~\ref{fig:xi_r}, we showed how the ionization parameter $\xi$ declines at larger radii.   In Figure~1 of \citet{Nagele2026arXiv260103349N}, we showed that the inner rings for both $\mdot = 0.01$ and $\mdot = 0.1$ have $\tau < 10$, and when $\mdot = 0.01$, optical depth of 10 is rarely reached.  In addition, there are significant vertical density gradients essentially everywhere inside the photosphere except close to the midplane.

If the emitting region is limited to a few $r_g$, then one might well think that nearly all that region has the same surface density. However, if the slope $q$ of the emissivity's radial profile is large and $\xi$ is constant, $\Sigma$, and the associated near-photosphere density, \textit{must change} by $\sim q$ orders of magnitude over this few $r_g$ (Eq. \ref{eq:xi}).

In the following subsection we demonstrate how these assumptions closed off access to mechanisms by which the observed Fe~K$\alpha$ profiles could be produced by different means than emerge from traditional phenomenological modeling.

\subsection{Different ionization states of Fe contributing to a single observed line}

\begin{figure}
\includegraphics[width=\linewidth]{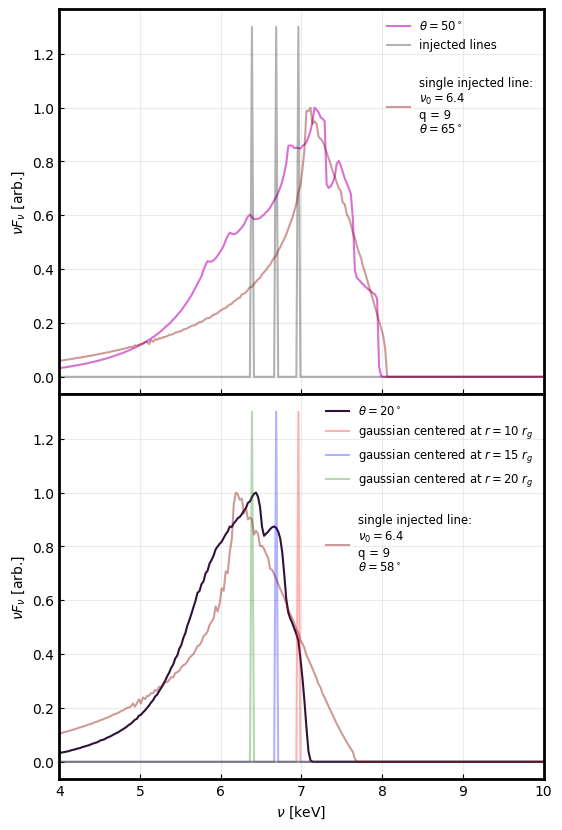}
    \caption{Normalized line profiles for multiple injected lines, at 6.4, 6.7, and 6.97 keV (grey lines). In these figures, we use \texttt{LineAART} with $q=2$, and the radial coverage extends from the ISCO to $r=100\;r_g$. Both panels also have a light red line comparing the profile to a single injected line with $q=9$ (cf. Fig. \ref{fig:Lineart_single}). In the top panel, the three lines are injected everywhere in the disk, and the $\theta=50^\circ$ viewing angle is shown. In the bottom panel, we show a less inclined viewing angle ($\theta=20^\circ$) and the three lines are injected as normal distributions in the radial coordinate, centered around $r=10,15,20\;r_g$ for 6.4, 6.7, and 6.97 respectively. Each distribution has a deviation of $10\;r_g$.  }
    \label{fig:three_lines}
\end{figure}

In Sec.~\ref{results:obs_flux}, we showed that multiple narrow lines emerge from the disk (Fig.~\ref{fig:Ldisk_observed}), particularly the He-like and H-like Fe~K$\alpha$ lines at 6.7 and 6.97~keV.  Because Doppler boosts, gravitational redshift, and scattering by hot electrons can shift the energy of a given line by $30\%$ (at $r=10\;r_g$, Fig. \ref{fig:double_horned}), lines whose rest-frame energies are separated by $< 20\%$ can all contribute to a single observed broad line (Fig. \ref{fig:Ldisk_sightlines}).
Thus, some of the broadness of the observed line is likely due to the emission of multiple narrow lines, either from multiple ionization states of a single element or from multiple elements. Combining multiple lines with Doppler shifts, etc. can create a line profile that appears to be the result of a single line broadened by larger Doppler shifts.
A similar observation was recently made by \citet{Biswas2025arXiv251103575B}, specifically in the context of warm coronae around an AGN accretion disk.

We demonstrate this phenomenon explicitly using \texttt{LineAART} profiles for a $q=2$ case (having constant $dL/dr$) in which the disk extends from the ISCO out to $r=100\;r_g$ (Fig. \ref{fig:three_lines}). In the top panel, we have injected three narrow lines at each location on this disk. When observed from $\theta=50^\circ$, each of these individual profiles shows a pronounced double horn pattern (Fig. \ref{fig:Lineart_single}, third panel). When observing the sum of the three lines, however, the total line is broader and the red and blue horns are 
nearly erased. In this idealized setting, the horns can be seen, but in a real observation they would be difficult to pick out over the noise and the power law continuum. 

In the bottom panel, we perform a similar exercise for a less inclined viewing angle ($\theta=20^\circ$). In this case, however, we localize the emission from the three lines to different parts of the disk,
placing the more highly-ionized Fe closer to the black hole (see Fig.~\ref{fig:xi_r}).
From a more face-on view, the Doppler effects are less important, but the difference in gravitational redshift between small radii and large cancels the energy shift due to different ionization states.
As a result, all three lines are observed at $\sim 6.4-6.7$~keV, mimicking a broad line from a single emission line confined to the region near the ISCO.

We demonstrate this last point by over-plotting $q=9$ line profiles from Fig. \ref{fig:Lineart_single}. The profiles from multiple lines are just as broad, if not broader, than the extreme $q=9$ profiles. Clearly, the feasibility of this multi-line merger effect depends strongly on the radial distribution of the different ionization stages, i.e., the radial distributions of $\xi$, $\Gamma$, and $\Sigma$.  In this idealized setting, we can easily see the differences between the two curves. In a more realistic setting, inhomogeneities in the surface brightness, Compton broadening, observational noise, and an uncertain continuum level all complicate the fitting procedure, and profiles of such similar shapes could not easily be distinguished. Even with the most advanced observing facilities, unfolding a single iron line into its rest frame components
involves ambiguities difficult to resolve.

\subsection{Non-uniqueness of Fe K$\alpha$ profiles: multiple ionization states, multiple elements, different surface brightness profiles, different inclinations}
\label{sec:non_uniqueness}

\begin{figure*}
\includegraphics[width=\linewidth]{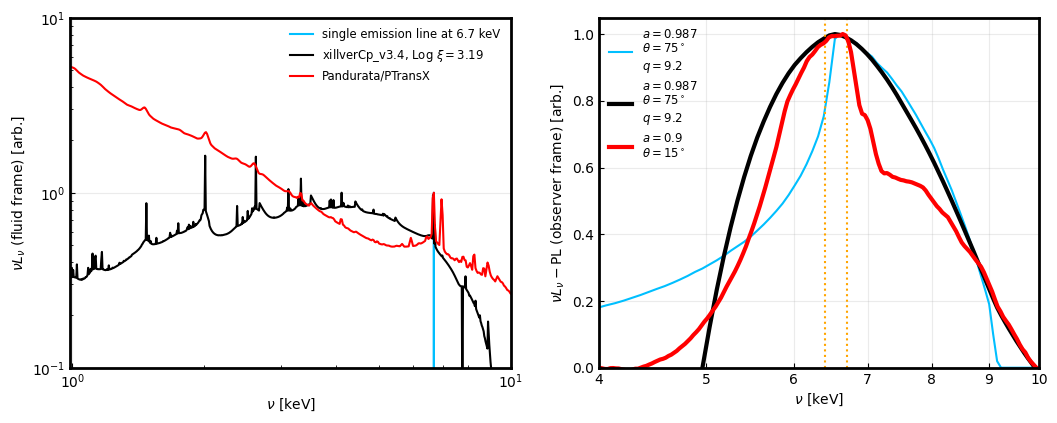}
    \caption{A comparison between a \texttt{xillver} spectrum and the \texttt{Pandurata/PTransX} disk spectrum in the context of observations of MAXI J1535-571 \citep{Xu2018ApJ...852L..34X}. The left panel shows the fluid frame spectrum of the indicated \texttt{xillver} model (black line) as well as the \texttt{Pandurata/PTransX} disk luminosity (red line). The blue line shows a single narrow emission line at 6.7 keV. The right panel shows the observed profile subtracted from a power law for the parameters indicated in the legend; the colors correspond to the physical cases listed in the left panel's legend. The yellow dotted lines mark the energies of neutral and He-like Fe~K$\alpha$, 6.4 and 6.7 keV, respectively.  For the narrow emission line, no power law is subtracted because there is no continuum emission.}
    \label{fig:comparison_profile}
\end{figure*}

The previous subsection is a specific example of a broader trend illuminated by our work: a single Fe~K$\alpha$ profile can result from 
multiple physical models. We show this explicitly for the case of NuSTAR observations of MAXI J1535-571 \citep{Xu2018ApJ...852L..34X}. The authors fit these observations with  with the \texttt{xillverCp} model \citep{Garcia2010ApJ...718..695G,Garcia2014ApJ...782...76G}, finding $a>0.987$, $q > 9.2$, $\log\xi=3.19$ and $\theta=75^{\circ} \,^{+2}_{-4}$ (model 4). We reproduce such a model in Fig. \ref{fig:comparison_profile}, and compare it to an iron line profile from this paper. The left panel shows the disk luminosity in the fluid frame for the $M=10\;\msun$, $\mdot=0.01$ model (Figs. \ref{fig:Ldisk_M}, \ref{fig:Ldisk_observed}), as well as an \texttt{xillverCp} spectrum with $\log \xi=3.19$, the value found by \citet{Xu2018ApJ...852L..34X}. These two fluid frame spectra in the Fe K$\alpha$ region are incredibly different, with this paper's dominated by a Compton-broadened iron line with narrow cores at 6.7 and 6.97 keV, while the \texttt{xillverCp} spectrum has a single narrow iron line at $6.7$ keV. The \texttt{xillverCp} spectrum also has a single iron absorption feature at $7.7$ keV while the \texttt{Pandurata/PTransX} is dominated by nickel emission lines in this energy range, due to the higher average ionization state (Fig. \ref{fig:xi_r}, see also Appendix B for a discussion of absorption features in fluid frame spectra). 

The right panel of Fig. \ref{fig:comparison_profile} shows the observed profile of the Fe K$\alpha$ line originating from the fluid frame spectrum in the left panel. For the \texttt{xillverCp} model, we have used \texttt{LineAArt} to account for the relativistic effects, as in Fig. \ref{fig:Lineart_total}. We use the following parameters, motivated by model 4 of \citet{Xu2018ApJ...852L..34X}: $a=0.987$, $q=9.2$, $r_{\rm in} = r_{\rm ISCO}$, $r_{\rm out}=10\;r_g$, $\theta=75^\circ$. The light blue curve (corresponding to a single emission line at 6.7~keV) reproduces the blue wing of the black curve, but not the red wing of the black curve. The red wing of the black curve has contributions from lower energy emission lines and the continuum. Once it is subtracted from a power law, the profile appears symmetric even though the original single line profile was notably asymmetric.

The red curve in Fig. \ref{fig:comparison_profile} shows the \texttt{Pandurata/PTransX} profile (as in Fig. \ref{fig:line_with_scattering}) for $\theta=15^\circ$. The two profiles are not identical, but are both good fits to the data \citep[Fig. 2 of ][]{Xu2018ApJ...852L..34X}. Interestingly, our model reproduces the notch at 7 keV seen in the observed system, and this is due to the emission lines at $\nu>7$ keV (left panel of Fig. \ref{fig:comparison_profile}), which mostly come from nickel. It is worth pointing out that our models have three parameters (mass, spin, and accretion rate) only one of which varies in this paper, so we do not necessarily expect good fits to any observational data. The fact that we obtain reasonable agreement with some observations strongly suggests that our approach is robust.

The degeneracy we have highlighted in this subsection is not alone. It has been known for some time that there is a bifurcation between low and high ionization parameter, with both producing reasonably good fits to the same data \citep{Bonson2016MNRAS.458.1927B}. The effects of absorption have also been shown to play a confusing role in BH parameter inference via X-ray reflection spectroscopy \citep{Miller2008A&A...483..437M,Kammoun2018A&A...614A..44K}. This is all on top of the fact that the majority of the parameter space is yet unexplored. The assumptions of constant $\xi$\footnote{Or the equally limiting assumption of computing $\xi$ from a lamppost corona.}$,\Gamma,$ and $\Sigma$ mean that there are entire classes of iron line profile fits which have not yet been considered.

\subsection{Simulations can guide us through the parameter space}
\label{sec:guide}

\begin{figure}[t!]
\includegraphics[width=\linewidth]{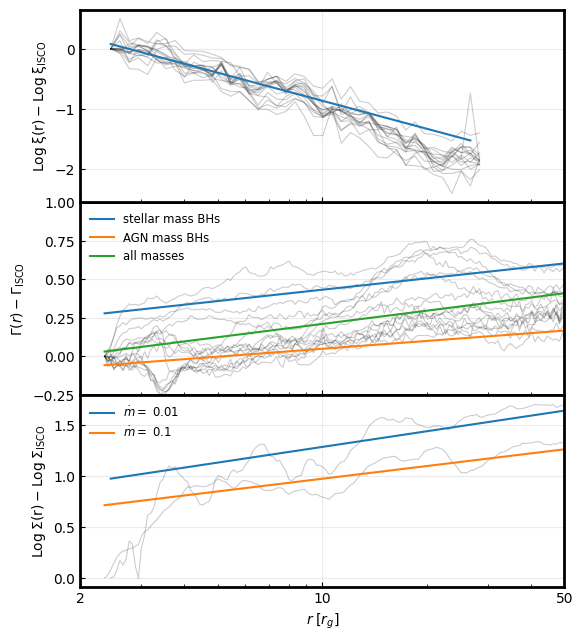}
    \caption{$\xi, \Gamma, $ and $\Sigma$ from the models in this paper (Figs. \ref{fig:Gamma_r}, \ref{fig:xi_r}), normalized to their values at the ISCO (grey curves). The colored curves are power law fits, noted in Table \ref{tab:slopes}.}
    \label{fig:power_law_slopes}
\end{figure}

Our calculations can help break the degeneracies of numerous non-unique models.  In particular, they point to specific radial dependences for the three basic quantities: $\xi$, $\Gamma$, and $\Sigma$.  Imposing those constraints can substitute for the assumptions of spatial homogeneity.
These radial dependencies can be modeled as power laws, which, for each of the three variables, reproduce our detailed calculations quite well (Fig. \ref{fig:power_law_slopes}). The slope of these power laws go roughly as:

\begin{equation}
\label{Eq:xi_r}
    \xi (r) = \xi_{\rm ISCO} \bigg(\frac{r}{r_{\rm ISCO}}\bigg)^{-1.5},
\end{equation}
\begin{equation}
\label{Eq:Gamma_r}
    \Gamma(r) = \Gamma_{\rm ISCO}  +  0.3 \log_{10} \bigg(\frac{r}{r_{\rm ISCO}}\bigg),
\end{equation}
\begin{equation}
\label{Eq:Sigma_r}
    \Sigma (r) = \Sigma_{\rm ISCO} \bigg(\frac{r}{r_{\rm ISCO}}\bigg)^{0.5}.
\end{equation}
Tab. \ref{tab:slopes} contains the details of the fits from Fig. \ref{fig:power_law_slopes}.

We have emphasized several times that the ionization parameter is not constant with radius, and that this assumption is a weakness in the approach of most phenomenological modelers. The fact that the radial dependence of $\xi$ can be modeled means that this dependence should be incorporated into phenomenological models. The further fact that the radial dependence can be modeled as a power law means that a {\it specific} $\xi(r)$ can be incorporated into phenomenological models at \textit{no extra computational cost,} with $\xi_{\rm ISCO}$ replacing $\xi$ in the fitting procedure (Eq. \ref{Eq:xi_r}). We note that $\xi(r)$ is well modeled by a power law not only by the sixteen models in this paper, but also by the seven models from \citet{Liu+2025} (Fig. \ref{fig:xi_paper_0}). 
 
$\Gamma$ varies by up to 0.8 in our simulation domain (Fig. \ref{fig:Gamma_r}), demonstrating that the softening of the incident flux with increasing radius also needs to be taken into account by fitting procedures. Like $\xi(r)$, $\Gamma (r)$ is well modeled by a power law (Fig. \ref{fig:power_law_slopes}). Thus, as with the ionization parameter, a variable photon index can be implemented into phenomenological fitting models with no extra computational cost, by replacing the parameter $\Gamma$ with $\Gamma_{\rm ISCO}$ (Eq. \ref{Eq:Gamma_r}).

The power law slopes of $\xi(r)$ and $\Gamma(r)$ (Tab. \ref{tab:slopes}) may change with different black hole parameters (e.g. spin), improvements in GRMHD simulations (e.g. thermodynamics), or improvements in post-processing (e.g. pairs or non-thermal electrons). We encourage modelers to therefore try out different values of these slopes, with the values found in this paper serving as starting points rather than the final word. Whichever values are used, however, we expect the black hole parameter inference to be quite different from the case of constant $\xi, \Gamma$ for the reasons enumerated earlier in the discussion.

$\Sigma(r)$ is a more challenging quantity to integrate into phenomenological models. Although these models often fit for an absorption column density, this density is somewhat removed from the physical density in the \texttt{xillver} (for example) slabs. This is because the absorption column density is measuring the amount of absorption from the photosphere to the observer, but this value does not yield information about the location of the photosphere. A disk density could, in principle, be inferred from the black hole mass and inferred accretion rate, yielding an approximate optical depth of the disk. But even if this value could be deduced, it would still take some work to implement it into the fitting procedure. \texttt{xillver} assumes that $\tau>>10$ while in many systems the optical depth is likely less than 10 at some radii (for instance, near the ISCO, see Fig. \ref{fig:power_law_slopes}). One could envision incorporating this effect into \texttt{xillver}, but it would require additional slabs to be run for a grid of $\tau$s in the range $\tau \in (0,10)$. In addition, these slabs would need to have flux incident on both sides of the slab (similar to the whole slabs in our simulations), as opposed to the current \texttt{xillver} setup which is semi-infinite.

Even if the correct optical depth were to be used, however, the precise value of the density depends on the scale height. There is also a physically grounded expectation that the density should increase towards the midplane instead of being constant in the vertical direction. For a discussion of the effects of treating slabs as having constant density, see Appendix A.

\begin{table}
\centering
\sisetup{scientific-notation=true, round-mode=places, round-precision=2}
\begin{tblr}{
colspec={cccc}
}  Variable & Description & Slope & Average ISCO value  \\ \hline
Log $\xi$ & --- & -1.6  &  5.4 \\ \hline
$\Gamma$ & stellar mass BHs & 0.24  &  1.6 \\
$\Gamma$ & AGN mass BHs & 0.17  &  1.4 \\
$\Gamma$ & all masses & 0.29  &  1.8 \\ \hline
Log $\Sigma$ & $\mdot=0.01$ & 0.51  &  -0.19 \\ 
Log $\Sigma$ & $\mdot=0.1$ & 0.41  &  1.7 \\ \hline
\end{tblr}
\caption{Power law slopes of several variables as they vary with radius, as illustrated in Fig.  \ref{fig:power_law_slopes}. The last column is the mean of the log values of each model which went into a particular fit. } \label{tab:slopes}
\end{table}

\subsection{Black hole spin}
\label{sec:spin}

One of the primary goals of X-ray reflection spectroscopy is to measure the spin parameter of black holes. As we have outlined, this effort usually attributes the majority of the width of broad Fe K$\alpha$ emission lines to arising from severe relativistic effects near the black hole. This method often infers extremely rapid rotation \citep{Jiang2019MNRAS.489.3436J,Draghis2025ApJ...989..227D}, as only rapidly rotating black holes can get enough iron ions close enough to experience the observed amount of broadening. 

In this paper, we have demonstrated that broad Fe K$\alpha$ emission lines can arise from modest relativistic shifts at much larger radii, in combination with Compton broadening in the disk atmosphere and multiple ionization states caused by spatial inhomogeneities in both matter and radiation. This demonstration suggests that many other configurations can also produce broad Fe K$\alpha$ emission lines through similar mechanisms (Sec. \ref{sec:non_uniqueness}), sowing serious doubt on the robustness of previous spin measurements.

\section{Conclusions} \label{sec:conclusions}

In this paper, we have performed high resolution radiation transfer post-processing of GRMHD simulations of black hole accretion. We have described the radiation field within and at the surface of the accretion disk, and the spectra seen by distant observers. We have focused particularly on the Fe K$\alpha$ line, showing how the broadness arises from a combination of Compton broadening within the disk, emission from multiple ionization stages, and relativistic effects. The emission line profiles we find are similar in shape and equivalent width to many observed iron lines. This work represents the first time GRMHD simulation data have been post-processed to produce high resolution spectra on the basis of thermal and ionization balance for a wide range of black hole masses. 

We have shown that spatial gradients in $\xi$, $\Gamma$, and $\Sigma$ have major impact on the emitted line profiles, although these are generally set to zero in nearly all phenomenological modeling.  In particular, a gradient in $\xi$ can lead to sufficient blending of H-like, He-like, and neutral Fe~K$\alpha$ lines that broad K$\alpha$ profiles can be produced even when most of the luminosity is radiated at $r \gtrsim 10 r_g $.

We have raised nearly as many questions as we have answered, however. Why is the $\mdot=0.1$ case so highly ionized relative to $\mdot=0.01$ and is this a real or artificial result? Where in the disk does the iron line luminosity peak? Fig. \ref{fig:surface_lines} suggests that it is outside $r=30\;r_g$, but how far outside? Will lower ionization states contribute more at these larger radii? Are the power law slopes of the ionization parameter in Figs. \ref{fig:xi_r}, \ref{fig:xi_paper_0} universal or are they limited to spin parameter 0.9?

In future work, we plan to answer these questions by exploring a wider parameter range and by adding additional physics to see what effect each new mechanism has on the predicted spectrum.  Examples include: GRMHD simulations with radiation pressure forces which are necessary for simulating the correct vertical density profile of the accretion disk \citep{Zhang2025ApJ...995...26Z}; the departure of the electron temperature from the ion temperature, which is thought to be small, but not negligibly so in this regime \citep{Kinch_2020}; the assumption of thermal electron distributions, which recent particle-in-cell simulations suggest may be augmented by a hard power-law tail in the $\sim$MeV range \citep{Groselj2026arXiv260100518G}; pair physics which could be included in either the GRMHD simulation or the post-processing; the effects of disk winds, which are thought to be important for understanding the spectra of many systems \citep[e.g.][]{Pinto2016Natur.533...64P}. It is possible that the inclusion of additional physics could alter some of the specific results of our calculation; indeed, contrasting how well codes with different physics predict observables (e.g., continuum shapes, Fe~K$\alpha$ profiles, etc.) should be a valuable pointer to the mechanisms most important for determining those observables.

We highlight the above questions not only to serve as guideposts for future work, but also to remind the reader that these questions can now be answered with radiation transfer post-processing. In this paper we have demonstrated the ability to compute--- from physical principles and an atomic database--- (amongst other elements) Fe K$\alpha$ emission line profiles directly from GRMHD simulations. This computation allows us to better understand the origin of accretion disk emission lines and serves as another probe with which to interrogate the fidelity of GRMHD simulations to nature.

\begin{center}  \label{Sec:ack}
    \textbf{Acknowledgments}
\end{center}

We thank Scott Noble, the author of \texttt{HARM3D}, for use of his code and for much help in running it. We thank Tim Kallman for help with \texttt{XSTAR} reaction rates. 
This work was partially supported by NSF Grant AST-2009260 and NASA TCAN grant 80NSSC24K0100.

\bibliography{references.bib}{}
\bibliographystyle{aasjournal}

\begin{center}  \label{Sec:ack}
    \textbf{Appendix A: Comparison of \texttt{PTransX} to \texttt{xillver}}
\end{center}

\begin{figure*}
\includegraphics[width=\linewidth]{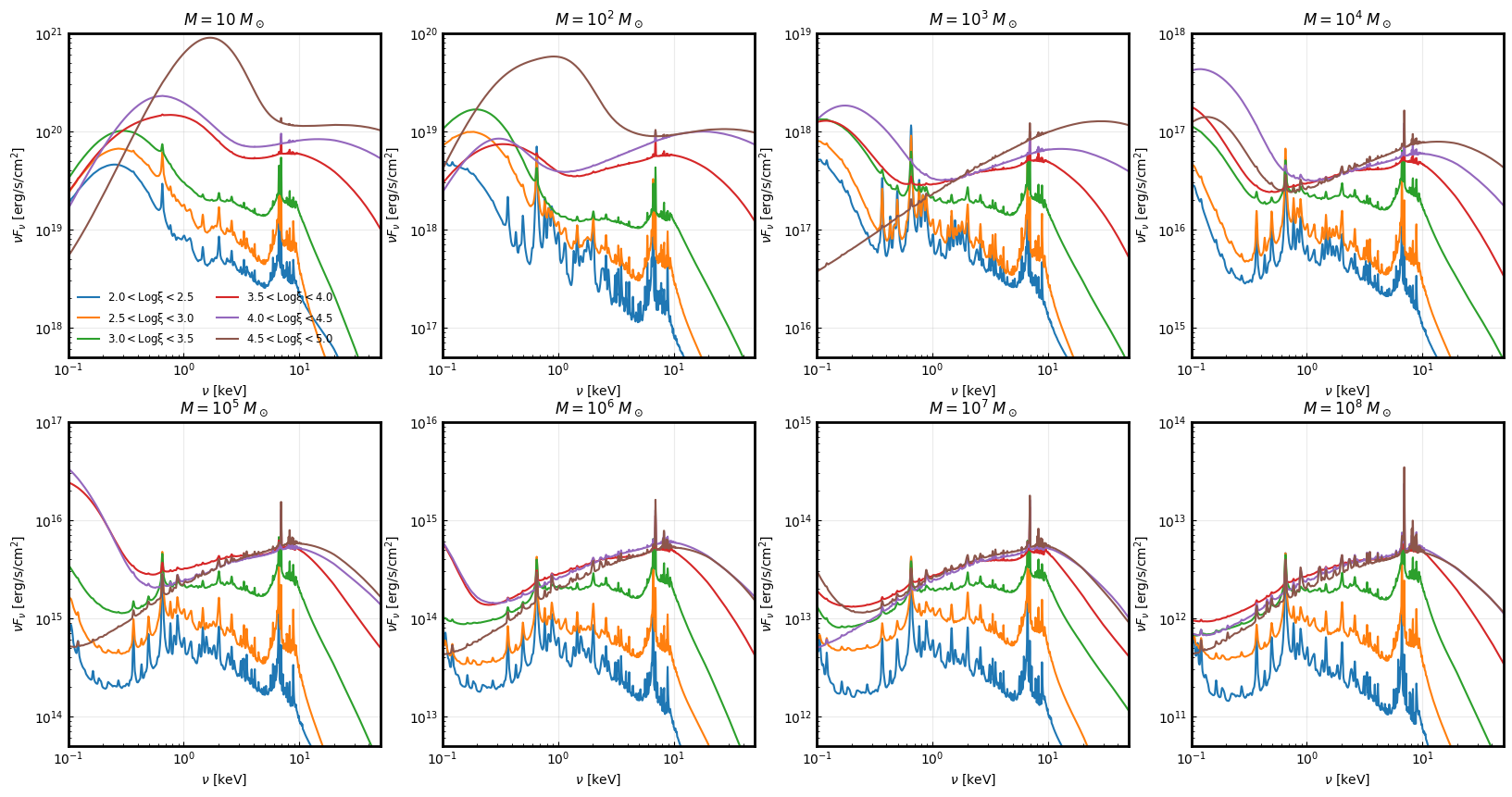}
    \caption{Averaged photospheric outgoing flux sorted into bins of Log $\xi$. Log $\xi$ follows a decreasing trend at larger radii (Fig. \ref{fig:xi_r}), so the higher ionization parameter bins are weighted towards the inner radii and vice versa. Each panel shows a different black hole mass for the $\mdot=0.01$ case. }
    \label{fig:flux_xi_sorted}
\end{figure*}

\begin{figure*}
\includegraphics[width=\linewidth]{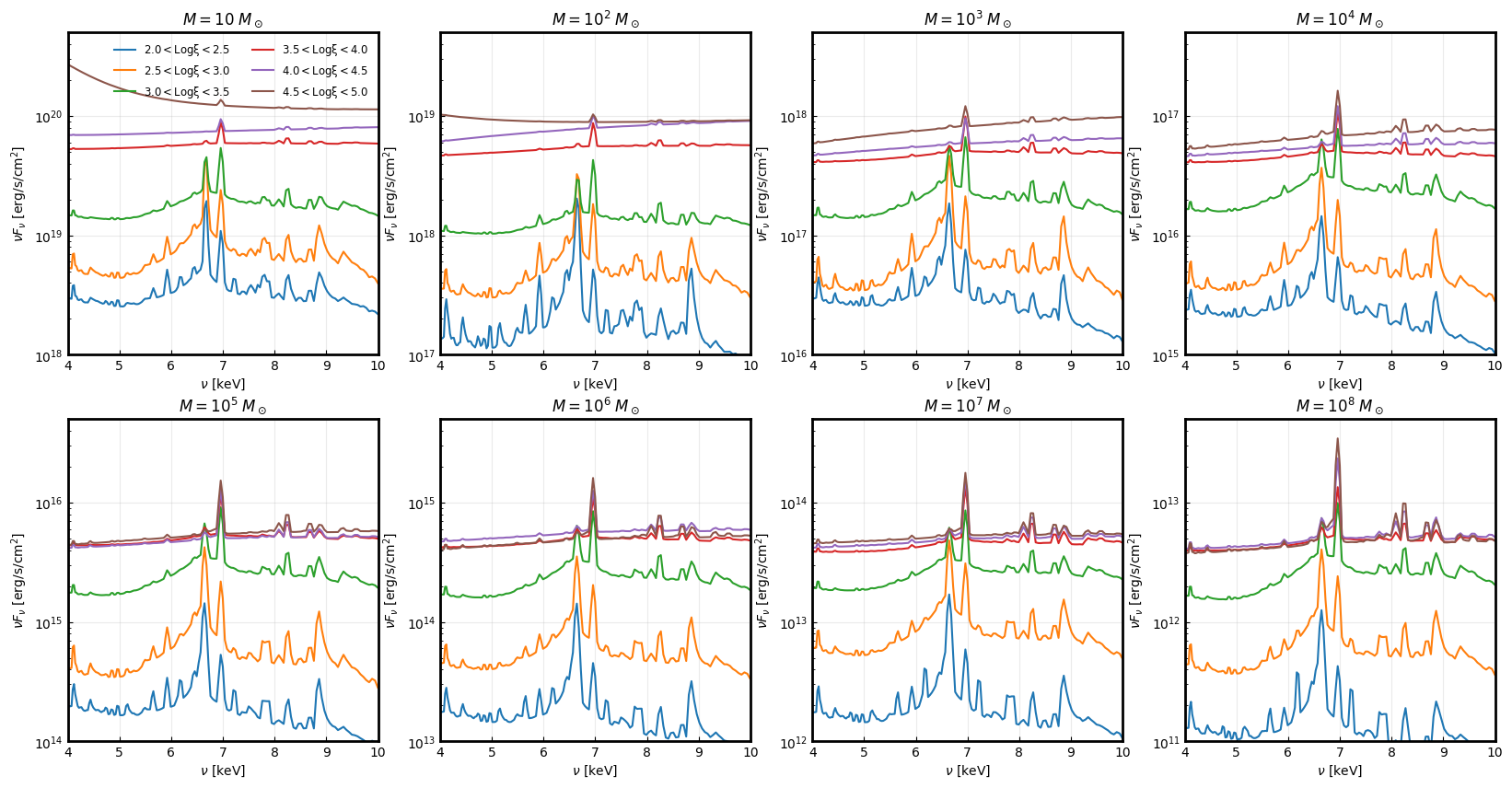}
    \caption{Same as Fig. \ref{fig:flux_xi_sorted}, but zoomed in on the Fe K$$}
    \label{fig:flux_xi_sorted_zoom}
\end{figure*}

Figs. \ref{fig:flux_xi_sorted}, \ref{fig:flux_xi_sorted_zoom} show the outgoing photospheric flux for all slabs for all of the $\mdot=0.01$ models (each panel shows a different mass). The fluxes are averaged after being binned in ionization parameter (bins over half a decade). Fig. \ref{fig:flux_xi_sorted} shows the full energy range while Fig. \ref{fig:flux_xi_sorted_zoom} shows the Fe k$\alpha$ region. These figures enable a very rough comparison to made with the fluid frame fluxes computed by phenomenological models. In particular, see Fig. 2 of \citet{Garcia2010ApJ...718..695G}, Fig. 5 of \citet{Garcia2013ApJ...768..146G}, Fig. 2 of \citet{Garcia2014ApJ...782...76G}, Fig. 5 of \citet{Garcia2020ApJ...897...67G}, and Fig. 1 of \citet{Huang2025arXiv251212728H}.

The main difference between the fluxes of different masses (different panels of Figs. \ref{fig:flux_xi_sorted}, \ref{fig:flux_xi_sorted_zoom}) and the fluxes of the phenomenological models with corresponding values of $\xi$, is not the shape of the emission lines, but the level of the continuum. In Fig. \ref{fig:flux_xi_sorted}, the spectrum has a prominent thermal peak at 0.2-2 keV for $M=10\;\msun$, and this peak shifts to lower energy as the mass increases \citep{Nagele2026arXiv260103349N}. In addition, the slope of the continuum in the Fe K$\alpha$ region changes with $\xi$. Further out in the disc, $\Gamma$ is usually softer (Fig. \ref{fig:Gamma_r}) and this can be seen in the softening of the continuum as $\xi$ decreases. Each particular slab in our simulation has a value of $\Gamma$ (or sometimes two, if it is a whole slab which has flux incident on both sides), and many also have thermal cores with a particular temperature. Thus each individual slab could be mapped to a particular \texttt{xillver} slab. The nuance that our post-processing setup allows is the particular ratio of all of these quantities at each specific location on the disc surface. We also note that many of the highest $\xi$ slabs are located near the ISCO and are thus thin ($\tau_{\rm slab}\ll 10$). These slabs do not have an analog in the phenomenological tables because those models assume a constant optical depth (usually $\tau_{\rm slab}=10$).

Another difference between the \texttt{PTransX} slabs and the \texttt{xillver} slabs is that \texttt{PTransX} uses a vertical density profile taken from a GRMHD simulation whereas \texttt{xillver} assumes a constant vertical density. This has two relevant consequences: the ratio of emission line to absorption line features and the strength of Compton broadening. In a constant density disk, one would expect a higher ratio of absorption features to emission line features because there is more mass directly adjacent to the photosphere, and it is the material in this region from which absorption features primarily arise. Indeed, our models do not show many absorption features and it will be interesting to see how this dynamic changes as we go to larger radius and correspondingly smaller $\xi$ (cf. Fig. 2 of \citealt{Garcia2010ApJ...718..695G}). Also, in a constant density disk, one expects less Compton broadening because the material directly adjacent to the photosphere will be less highly ionized for a given ionization parameter. For example, at Log $\xi = 3.5$, a constant density disk may have H-like ions at modest optical depths of 1-2, whereas a disk with an increasing density away from the photosphere will be more highly ionized in the vicinity of the photosphere. Thus, H-like ions will appear at somewhat larger optical depths, e.g. 2-3. This difference can be equivalently understood as the two types of slabs having different photospheric densities relative to the average slab density.

\begin{center}  \label{Sec:ack}
    \textbf{Appendix B: Fe K$\alpha$ profile gallery}
\end{center}

\begin{figure*}
\includegraphics[width=\linewidth]{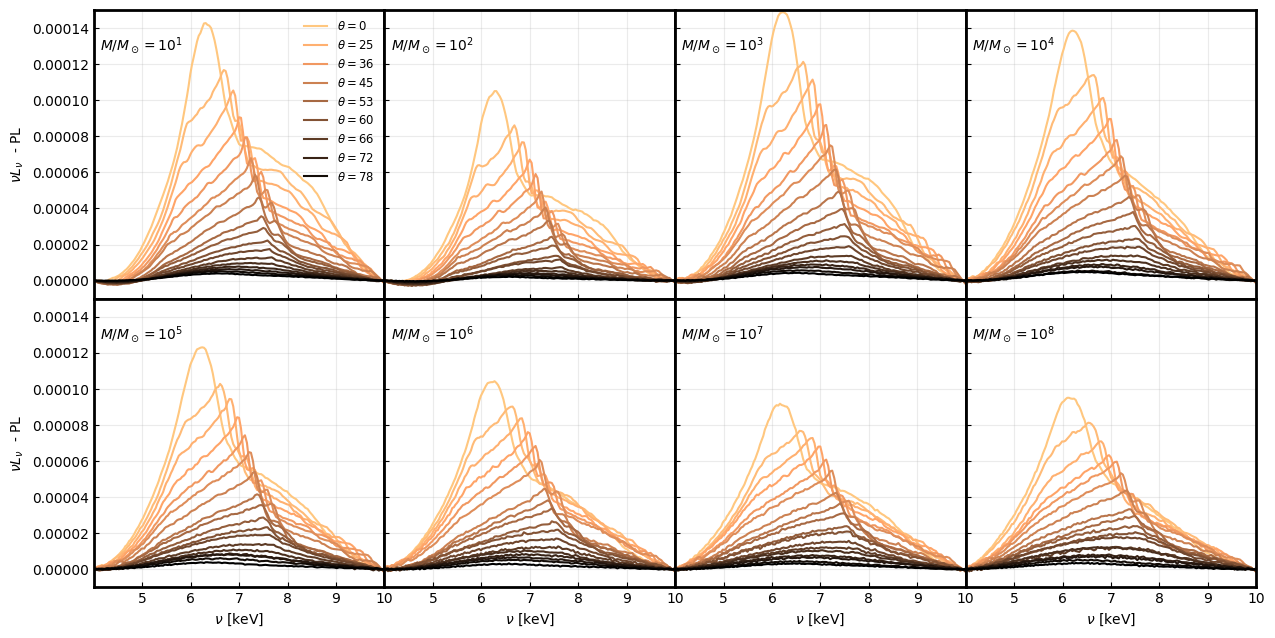}
    \caption{Same as dark lines in Fig. \ref{fig:line_with_scattering}, but for all masses and all northern hemisphere angles. Every second line is labeled. }
    \label{fig:all_ms_thetas_northern}
\end{figure*}

\begin{figure*}
\includegraphics[width=\linewidth]{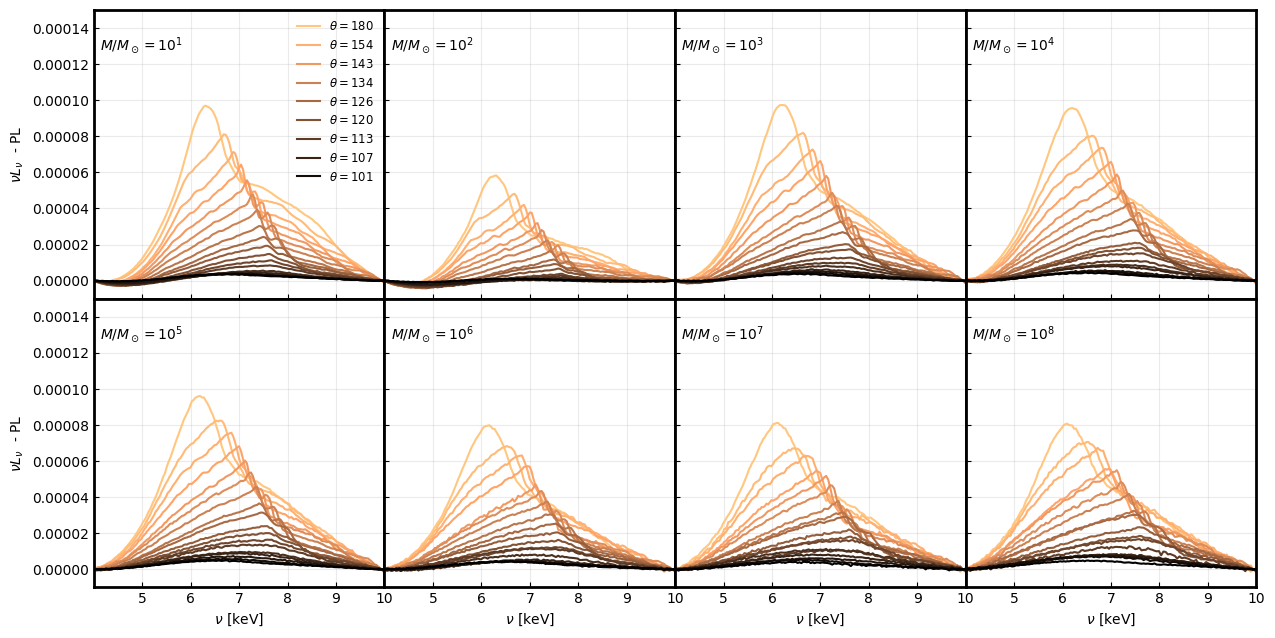}
    \caption{Same as Fig. \ref{fig:all_ms_thetas_northern}, but for the southern hemisphere. }
    \label{fig:all_ms_thetas_southern}
\end{figure*}

Figs. \ref{fig:all_ms_thetas_northern}, \ref{fig:all_ms_thetas_southern} show all of the Fe K$\alpha$ profiles for this paper (cf. dark lines in Fig. \ref{fig:line_with_scattering}).

\end{document}